\documentclass[
journal=jpcafh,
manuscript=article,
maxauthors=10,
etalmode=truncate
]{achemso}

\usepackage{chemmacros}   
\usepackage[T1]{fontenc}
\usepackage{easyReview}
\usepackage{multicol}     
\usepackage{multirow}     
\usepackage{appendix}

\usepackage{amsmath}      
\usepackage{cases}        

\usepackage[pdftex]{hyperref}
\usepackage{cleveref}                  
\mciteErrorOnUnknownfalse 

\usepackage{ctable}       
\usepackage{colortbl}     
\usepackage[para]{threeparttable} 

\usepackage{subfig} 

\usepackage{doi}          

\makeatletter
\newcommand{\setlabel}[1]{\edef\@currentlabel{#1}\label}
\makeatother

\newcommand{\greymidrule}{\arrayrulecolor{black!10}\midrule\arrayrulecolor{black}}

\author{\'{E}lio Pereira}
\affiliation{Instituto de Plasmas e Fus\~{a}o Nuclear, Instituto Superior T\'{e}cnico, Universidade de Lisboa, Av. Rovisco Pais 1, Lisboa, 1049-001, Portugal}
\author{Jorge Loureiro}
\affiliation{Instituto de Plasmas e Fus\~{a}o Nuclear, Instituto Superior T\'{e}cnico, Universidade de Lisboa, Av. Rovisco Pais 1, Lisboa, 1049-001, Portugal}
\author{M\'{a}rio Lino da Silva}
\email{mlinodasilva@tecnico.ulisboa.pt}
\affiliation{Instituto de Plasmas e Fus\~{a}o Nuclear, Instituto Superior T\'{e}cnico, Universidade de Lisboa, Av. Rovisco Pais 1, Lisboa, 1049-001, Portugal}

\title[Vibronic State-Specific Modelling of High-Speed Nitrogen Shocked Flows. Part II: Shock Tube Simulations]{Vibronic State-Specific Modelling of High-Speed Nitrogen Shocked Flows. Part II: Shock Tube Simulations}


\begin{document}







\begin{abstract}
The conditions of thermochemical and radiative non-equilibrium attained in nitrogen shocked flows were quantified using a vibronic state-specific model. This model, being described in a companion paper, was implemented in Euler one-dimensional simulations for shots $19$, $20$ and $40$ of the EAST's $62^{\text{th}}$ campaign. It was found that the peak values of the instrumental radiative intensities were underestimated by one to two orders of magnitude, and sensitivity tests performed on several parameters of the simulations were not successful in getting a reasonable agreement. The shapes of the instrumental radiative intensities' profiles obtained in the low speed shot were correctly predicted, unlike the ones of the medium and high speed shots which revealed non-null plateaus proceeding or coalescing with peaks. These plateaus were not predicted at all. It is suspected that such discrepancies may have resulted from neglecting other shock tube related phenomena, as pointed out by other researchers in the literature: the absorption of radiation emitted by the driver gas and the EAST electric arc, and/or the conduction of heat due to downstream plasma being subjected to a stronger shock wave.
\end{abstract}

\section{Introduction}
\label{sec:Introduction}
\setlabel{Introduction}{sec:introduction}

The main objective of Spacecraft Design with respect to atmospheric entries is to devise a vehicle that can sustain the harsh conditions of the flight, descend in a stable and controllable  way, and decelerate smoothly so that landing can safely happen. Regarding the survival of the spacecraft against the harsh conditions, one should focus on the strong convective and radiative heats received by the vehicle. The distribution of the heat around the body needs to be well predicted for a correct estimation of the thickness of the thermal protection system (TPS). A TPS with an underestimated thickness would compromise the integrity of the vehicle due to the removal of a considerable part of the protective material by ablative processes. On the other hand, a TPS with an overestimated thickness would make the spacecraft unnecessarily heavier. And the higher the structural weight, the higher the required rate of ejected fuel to provide the same  acceleration, at launch and during manoeuvres in outer Space. To properly design the TPS, the interaction between the flow and the body, in a macroscopic and microscopic way needs to be understood. Such phenomena are dictated by the history of the elements of fluid that traverse the shock wave and arrive in the boundary layer of the body. In this work, post-shock flows that develop inside shock tubes were studied as an approximation to the ones that occur in atmospheric entries before the arrival of the elements of fluid in the boundary layer. In fact, the main objective of the work was to validate the developed vibronic-specific state-to-state model for pure nitrogen shocked flows, which was described in a companion paper \cite{epereira2021paperI}, by performing simulations of benchmark shots done in a shock tube whose results were available in the literature. Note that nitrogen is the main component of Earth's and Titan's atmosphere. By firstly neglecting the other components, one may focus on a single part of the entire model, and validate it through shock tube tests that consider pure nitrogen as test gas. One would then shift the focus to the other components in future works, validating the respective models and considering models for the interactions between the components, finally adding them to get the complete model.
The shock tube tests which were regarded in this work were the ones from the $62^{\text{th}}$ campaign of the Ames Electric Arc Shock Tube (EAST), performed in $2018$ by Brandis and Cruden \cite{brandis2018shock}. The curious reader may get to know more about this shock tube facility by reading the works of Sharma and Park \cite{sharma1990a,sharma1990b}. In the $62^{\text{th}}$ campaign, a driven gas of pure nitrogen (\ch{N2}) was considered. The tests comprised upstream (of the shock wave) flow speeds $u_{\infty}$ between approximately $6$ and $11\,\text{km}/\text{s}$, a static pressure of value $p_{\infty}=0.2\,\text{Torr}$, and a temperature of value $T_{\infty}=300\,\text{K}$. 

The shock tube tests were here simulated using the SPARK code \cite{lopez2016}.
The following sections present a set-up of the numerical simulations, their results, and a discussion on the discrepancies between these and experimental results.

\section{Theory}
\label{sec:theory}
\setlabel{Theory}{sec:theory}

In this section, the shock tube experiment, the variables measured by Brandis and Cruden in the experiment, as well as the fluid flow governing equations and their boundary conditions which were regarded in this work, will be described.

\subsection{The shock tube experiment}\label{subsec:apparatus}
\begin{figure}[H]
\centering
\centerline{\includegraphics[scale=1]{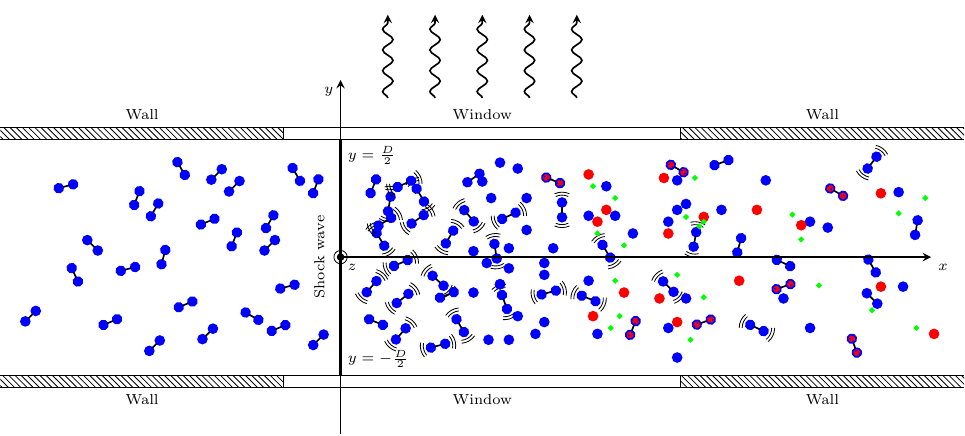}}
\caption{Upper view of a longitudinal cross-section of the shock tube, showing the measured radiative field.}
\label{fig:EAST_apparatus}
\end{figure}

\Cref{fig:EAST_apparatus} presents a simplistic scheme of the flow inside the Ames Electric Arc Shock Tube during a shot, introducing the working frame of reference.
The shock tube, which is made of stain-less steel \cite{bogdanoff2002}, has an inner diameter $D=10.16\,\text{cm}$ \cite{brandis2018shock}. In each shot, the driver gas is heated by an electric arc, and the diaphragm that separates the driver from the driven gas opens. This produces a pressure discontinuity, resulting in a shock wave that propagates in the driven gas. The radiation emitted by the driven gas in the test section of the shock tube is then measured through a window by four spectrometers, each one associated with an interval of wavelengths $\lambda$. In the present work, the considered wavelength intervals  were the ones of the vacuum ultra-violet radiation (VUV), with $\lambda\in[145,\,195]\,\text{nm}$, ultra-violet/visible radiation (``Blue''), with $\lambda\in[330,\,480]\,\text{nm}$, visible/near infra-red radiation (``Red''), with $\lambda\in[480,\,890]\,\text{nm}$, and infra-red radiation (IR), with $\lambda\in[890,\,1450]\,\text{nm}$. The post-shock flow may be considered to be stationary (not varying in time, but in space) when observed from a frame of reference that moves with the shock wave if this shock wave does not decelerate too much while travelling inside the shock tube (note, however, that such deceleration is indeed known to occur\cite{bogdanoff2002}). The frame of reference here mentioned is depicted in \Cref{fig:EAST_apparatus}, where its origin coincides with the point of intersection of the shock tube axis with the shock wave plane, its $x$-axis has the direction of the shock tube axis and points in the downstream direction, its $z$-axis points upwards, and its $y$-axis is such that $\vec{e}_x\times\vec{e}_y=\vec{e}_z$, pointing in a radial direction. It is also worthy to refer here that the flow may be regarded as unidimensional if the boundary layer does not grow too much.

\subsection{Measured variables}\label{subsec:measurements}

Brandis and Cruden measured the post-shock specific radiative intensity $I_\lambda$ associated with photons propagating in the radial direction of the tube $\vec{e}_y$, at the farthest radial point of the plasma, $y=D/2$, as depicted in \Cref{fig:EAST_apparatus}.
One may further quantify the specific radiative intensity through the so-called equation of radiative transfer \cite{vincenti1965introduction} for the propagation direction $\vec{e}_y$ disregarding variations in time (due to the hypothesis of stationarity), the dependence of the emission and absorption coefficients on $y$ and $z$ (due to hypothesis of unidimensionality), and induced emission, as well as initially assuming the medium to be optically thin and then introducing the concept of an escape factor \cite{NEQAIR96}, $\Lambda_{s,e,v}^{e',v'}\in[0,\,1]$. This escape factor corresponds to the fraction of the emitted photons due to spontaneous emissions from the vibronic level $(e,\,v)$ to $(e',\,v')$ which escape the system, or in other words, that are not absorbed. 
One has then
\begin{equation}
I_{\lambda}(x,\lambda)=\frac{hcD}{\lambda}\left[\sum_{\text{se}}\Lambda_{s,e,v}^{e',v'}\frac{A_{s,e,v}^{e',v'}}{4\pi}\,\phi_{\lambda,s,e,v}^{e',v'}(x,\lambda)\,n_{s,e,v}(x)\right]\text{ ,}
\label{eq:I_shock_tube_5}
\end{equation}
where the sum is done in all spontaneous emission processes. Note that in \eqref{eq:I_shock_tube_5}, $h$ is the Planck's constant, $c$ is th speed of light, $A_{s,e,v}^{e',v'}$ is the Einstein coefficient for spontaneous emission, $\phi_{\lambda,s,e,v}^{e',v'}$ is the line-shape factor and $n_{s,e,v}$ is the number density of $s$-th species particles in the $(e,\,v)$ vibronic level.

Among the data issued by Brandis and Cruden there are the instrumentally resolved radiative intensities $\hat{I}(x)$ and the instrumentally resolved non-equilibrium metrics $\hat{I}_\lambda^{\,\text{ne}}(\lambda)$ for each of the four intervals of wavelengths in a total of $42$ shots. The hat denotes instrumentally resolved quantities, in contrast to the real quantities. The former are the ones that are actually obtained in the experiment, being related to the latter through a transformation ``applied'' by the instruments. The instrumentally resolved radiative intensity associated with the $l$-th wavelength interval, with $l\in\{\text{VUV},\,\text{``Blue''},\,\text{``Red''},\,\text{IR}\}$, is given by an integration of the instrumentally resolved specific radiative intensity $\hat{I}_\lambda(x,\lambda)$ with respect to the wavelength $\lambda$ from $\lambda=\lambda_{\text{min}}^l$ to $\lambda=\lambda_{\text{max}}^l$:
\begin{equation}
\hat{I}^l(x)=\int_{\lambda_{\text{min}}^l}^{\lambda_{\text{max}}^l}\hat{I}_\lambda(x,\lambda)\,d\lambda\text{ .}
\label{eq:I_hat}
\end{equation}
And the instrumentally resolved non-equilibrium metric associated with the $l$-th interval of wavelengths is given by an integration of the instrumentally resolved specific radiative intensity $\hat{I}_\lambda(x,\lambda)$ with respect to the position $x$ from $x=x_{\text{min}}^l$ to $x=x_{\text{max}}^l$ - a region characterised by a strong thermodynamic non-equilibrium - divided by the shock tube inner diameter $D$:
\begin{equation}
\hat{I}_{\lambda}^{\text{ne},l}(\lambda)=\frac{1}{D}\int_{x_{\text{min}}^l}^{x_{\text{max}}^l}\hat{I}_\lambda(x,\lambda)\,dx\text{ .}
\label{eq:I_hat_ll_ne}
\end{equation}

One should note here that it is not possible to know the exact position of the shock wave through the spectra obtained in the shock tube experiments done by Brandis and Cruden for multiple reasons: the onset of the radiative field is associated with an increase of the number of excited species (which occurs at some distance downstream of the shock wave), and each spectrometer captures the spectra after some reaction time \cite{cruden2012c}. The $x$ axis that Brandis and Cruden work with may not correspond to the one depicted in Figure \ref{fig:EAST_apparatus}. The issued position values are actually with respect to a particular origin for each shot and interval of wavelengths which may not coincide with a point in the shock wave. Therefore, they will be denoted here as $\hat{x}$ and termed ``relative positions'' to distinguish them from the previously introduced ones, $x$, which are with respect to an origin that coincides with a point in the shock wave. The relative position of the shock wave for some shot and the $l$-th interval of wavelengths, say $\hat{x}_\text{sw}^l$, was defined in this work as the point where the instrumentally resolved radiative intensity starts to rise abruptly, as also considered by Cruden in his work \cite{cruden2012c}. The position $x$ is then given by $x=\hat{x}-\hat{x}_{\text{sw}}^l$. 
The wavelength integration limits $\lambda_\text{min}^l$ and $\lambda_\text{max}^l$, and relative position integration limits $\hat{x}_\text{min}^l$ and $\hat{x}_\text{max}^l$ are issued by Brandis and Cruden in their work.
The instrumentally resolved specific radiative intensity $\hat{I}_{\lambda}^l(x,\lambda)$ departs from the real one, $I_{\lambda}^l(x,\lambda)$, either spectrally and spatially. The non-ideality of the instrumental apparatus is such that the measured specific radiative intensity $\hat{I}_{\lambda}^l(x,\lambda)$ associated with some particular wavelength $\lambda$ and position $x$ is in fact the result of a distribution of the real radiative intensity on intervals of wavelengths and positions around the reference values. Or, mathematically
\begin{equation}
\hat{I}_{\lambda}(x,\lambda)=\int_{-\infty}^{\infty}\hat{\phi}^{\text{spa}}(x')\left[\int_{-\infty}^{\infty}\hat{\phi}^{\text{spe}}(\lambda')I_{\lambda}(x-x',\lambda-\lambda')\,d\lambda'\right]\,dx'\text{ ,}
\label{eq:spe_spa_convolution}
\end{equation}
with $\hat{\phi}^{\text{spe}}(\lambda')$ and $\hat{\phi}^{\text{spa}}(x')$ being the so-called instrument line-shape factor and spatial resolution function, respectively, which are issued by Brandis and Cruden in their work.

\subsection{Fluid flow governing equations}\label{subsec:fluid_flow}

As referred before, in this work it was assumed that the post-shock flow was unidimensional and stationary. Moreover, it was assumed that the transport phenomena (mass diffusion, viscosity and heat conduction) were negligible (the flow is then said to be of the Euler type) as well as the external body forces, and that the translational and rotational energy modes of the heavy particles were in equilibrium with each other (which is normally true immediately downstream of the shock wave \cite{park1990nonequilibrium}). The equations which were dealt with were then the ones of balance of the species' mass in their vibronic levels (due to the regarded vibronic-specific state-to-state model), momentum, total energy, and free electrons energy subjected to the above-mentioned assumptions:
\begin{subnumcases}{}
\frac{d c_{s,e,v}}{d x}=\frac{\dot{\omega}_{s,e,v}}{\rho u},\,\,\,\,\,\,\,\,\,\forall s,\,e\,\text{and }v\text{ ,}
\label{eq:mass_equation_ve_1D_final}
\\
\left(\frac{\rho u^2}{p}-1\right)\frac{du}{dx}+\frac{u}{p}\left[\left(\sum_{s\in\{\text{h}\}}\frac{p_s}{T_{\text{tr}_\text{h}}}\frac{dT_{\text{tr}_\text{h}}}{dx}\right)+\frac{p_\text{e}}{T_{\text{tr}_\text{e}}}\frac{dT_{\text{tr}_\text{e}}}{dx}\right]=-\frac{1}{p\rho}\left[\left(\sum_{s\in\{\text{h}\}} \frac{\dot{\omega}_s p_s}{c_s}\right)+\frac{\dot{\omega}_\text{e} p_\text{e}}{c_\text{e}}\right]\text{ ,}
\label{eq:momentum_equation_1D_final}
\\
\parbox{
\widthof{$
\displaystyle
=-\frac{\dot{\Omega}_{\text{rad}}+\left(\sum_s\dot{\Omega}_{s,\text{e}}^{\text{int}}\right)+\left[\sum_{s\in\{\text{h}\}}\dot{\omega}_s\left(h_s+\frac{1}{2}u^2\right)\right]+\left[\sum_{s\in\{\text{h}\},e,v}\left(\dot{\omega}_{s,e,v}-\frac{c_{s,e,v}}{c_s}\dot{\omega}_s\right)\frac{\epsilon_{s,\text{el-vib},e,v}}{m_s}\right]}{\rho u\left(\sum_{s\in\{\text{h}\}}c_sC_{p,s,\text{tr-rot}}\right)}\text{ ,}
$}}
{$
\displaystyle
\frac{d T_{\text{tr}_\text{h}}}{dx}+\frac{\left(\sum_{s\in\{\text{h}\}} c_s\right)u}{\sum_{s\in\{\text{h}\}}c_sC_{p,s,\text{tr-rot}}}\cdot\frac{du}{dx}=\\
=-\frac{\dot{\Omega}_{\text{rad}}+\left(\sum_s\dot{\Omega}_{s,\text{e}}^{\text{int}}\right)+\left[\sum_{s\in\{\text{h}\}}\dot{\omega}_s\left(h_s+\frac{1}{2}u^2\right)\right]+\left[\sum_{s\in\{\text{h}\},e,v}\left(\dot{\omega}_{s,e,v}-\frac{c_{s,e,v}}{c_s}\dot{\omega}_s\right)\frac{\epsilon_{s,\text{el-vib},e,v}}{m_s}\right]}{\rho u\left(\sum_{s\in\{\text{h}\}}c_sC_{p,s,\text{tr-rot}}\right)}\text{ ,}
$}
\label{eq:T_trh_rot_equation_1D}
\\
\frac{d T_{\text{tr}_\text{e}}}{d x}+\frac{u}{C_{p,\text{e}}}\frac{du}{dx}=\frac{\left(\sum_s\dot{\Omega}_{s,\text{e}}^{\text{int}}\right)-\dot{\omega}_\text{e}\left(h_\text{e}+\frac{1}{2}u^2\right)}{\rho u c_\text{e}C_{p,\text{e}}}\text{ .}
\label{eq:T_tre_equation_1D}
\end{subnumcases}
In the equation for the balance of the species' mass in their vibronic levels \eqref{eq:mass_equation_ve_1D_final}, $c_{s,e,v}$ is the mass fraction of $s$-th species particles in the $(e,\,v)$ vibronic level, $\dot{\omega}_{s,e,v}$ is the respective volumetric mass source term, $\rho$ is the mixture's mass density, and $u$ is the $x$-component of the mixture's flow velocity vector. In the equation for the balance of the mixture's momentum \eqref{eq:momentum_equation_1D_final}, $\{h\}$ denotes the set of heavy species, $p$ is the mixture's pressure, $p_s$ is the partial pressure associated with the $s$-th species particles, $T_{\text{tr}_\text{h}}$ is the heavy particle translational temperature, $T_{\text{tr}_\text{e}}$ is the free electron translational temperature, and the quantities $\dot{\omega}_\text{e}$, $p_{\text{e}}$ and $c_\text{s}$ are $\dot{\omega}_\text{s}$, $p_{\text{s}}$ and $c_\text{s}$ in which the $s$-th species corresponds to the free electron. In the equation for the balance of the total energy \eqref{eq:T_trh_rot_equation_1D}, $C_{p,s,\text{tr-rot}}$ is the translational-rotational specific heat at constant pressure of the $s$-th species particles, $\dot{\Omega}_\text{rad}$ is the volumetric radiative energy source term, $\sum_s \dot{\Omega}_{s,\text{e}}^{\text{int}}$ is the energy transferred per unit of time and volume from the inner particles to the inner free electrons of the element of fluid, $h_s$ is the specific enthalpy of the $s$-th species particles, $\epsilon_{s,\text{el-vib},e,v}$ is the sensible vibronic energy of a $s$-th species particle in its $(e,\,v)$ vibronic level, and $m_s$ is its mass. Finally, in the equation for the balance of the free electrons energy \eqref{eq:T_tre_equation_1D}, $C_{p,\text{e}}$ is the free electrons specific heat at constant pressure.


\subsubsection{Initial conditions}\label{sssec:init_cond}
Initial values need to be assigned to the unknowns of the problem. These are associated with the conditions of the post-shock flow immediately downstream of the shock wave - which will be labelled here by ``2''. It can be assumed that no chemical processes neither electronic or vibrational excitation occur throughout the shock wave \cite{park1990nonequilibrium}. Hence, $c_{s,v,e,2}=c_{s,v,e,\infty}$, where the symbol ``$\infty$'' denotes the upstream conditions. Particular attention should be given to the free electron translational temperature $T_{\text{tr}_\text{e}}$: since the free electrons are much lighter than the heavy particles, the excitation of their translation energy modes occurs in a different way, and it is not certain that an equilibration between them occurs immediately downstream of the shock wave. Also, the mole fraction of free electrons should be negligible throughout the shock wave, which may let us to regard the problem as a purely numerical one. Since no discussion about the topic was found in the literature, it was decided to regard the choice of other researchers, such as Kadochnikov and Arsentiev \cite{kadochnikov2020}, and consider the equality $T_{\text{tr}_\text{e},2}=T_{\text{tr}_\text{h},2}$. This equality implies that the  immediately freed electrons have the same translational temperature as the heavy particles to which they were initially bonded.

The low $T_{\infty}$ value (of $300\,\text{K}$) allows one to disregard the contribution of the vibrational and electronic energy modes to the particles energy \cite{anderson2006hypersonic} throughout the shock wave. This means that both mixture specific heat at constant volume $C_V$ and mixture specific heat at constant pressure $C_p$, from the upstream to the immediately downstream conditions, only depend on the translational and rotational energy modes of the particles. It is then possible to obtain the well-known Rankine-Hugoniot jump conditions, which allows one to compute the conditions immediately downstream of the shock wave from the ones upstream of it \cite{vincenti1965introduction}. 
%

\subsection{The line-shape factor}
\label{subsection:line_shape_factor}
In this work, the line-shape factor $\phi_{\lambda,s,e,v}^{e',v'}$ regarded in \eqref{eq:I_shock_tube_5} was considered to be the result of four contributions: Doppler, collisional, Stark, and resonance broadening. One may determine a line-shape factor for each isolated contribution, and then compute the global line-shape factor $\phi_{\lambda,s,e,v}^{e',v'}$ through a triple convolution \cite{griem1997}:
\begin{multline}
\phi_{\lambda,s,e,v}^{e',v'}(\vec{r},t,\lambda)=\int_{-\infty}^{\infty}\left(\phi_{\lambda,s,e,v}^{e',v'}\right)_{\text{D}}(\vec{r},t,\lambda_{s,e,v}^{e',v'}+\lambda')\left\{\int_{-\infty}^{\infty}\left(\phi_{\lambda,s,e,v}^{e',v'}\right)_{\text{col}}(\vec{r},t,\lambda_{s,e,v}^{e',v'}+\lambda'')\left[\vphantom{\int_{-\infty}^{\infty}}\right.\right.\\
\left.\left.\int_{-\infty}^{\infty}\left(\phi_{\lambda,s,e,v}^{e',v'}\right)_{\text{S}}(\vec{r},t,\lambda_{s,e,v}^{e',v'}+\lambda''')\cdot\left(\phi_{\lambda,s,e,v}^{e',v'}\right)_{\text{res}}(\vec{r},t,\lambda-\lambda'-\lambda''-\lambda''')\,d\lambda'''\right]\,d\lambda''\right\}\,d\lambda'\text{ .}
\label{eq:global_line_shape}
\end{multline}

The line-shape factor for isolated Doppler broadening $\left(\phi_{\lambda,s,e,v}^{e',v'}\right)_{\text{D}}$ associated with radiative transitions between the vibronic levels $(e,\,v)$ and $(e',\,v')$ can be shown to be approximately given by a Gaussian function \cite{griem1997} $G(\lambda,w_G,\lambda_{0,G})$ of half-width at half-maximum \cite{penner1959} $\left(w_{\lambda,s,e,v}^{e',v'}\right)_{\text{D}}=\lambda_{s,e,v}^{e',v'}\sqrt{[2\ln2\,k_BT_{\text{tr}_\text{h}}]/(m_sc^2)}=:w_G$, centred at $\lambda=\lambda_{s,e,v}^{e',v'}=:\lambda_{0,G}$, with $k_B$ being the Boltzmann constant, and $\lambda_{s,e,v}^{e',v'}$ the wavelength of the photon in the absence of line broadening and shift.

For the case of collisional broadening, the formula based on the electron theory of Lorentz and referred by Penner \cite{penner1959} was considered for the respective line-shape factor. This is approximately given by a Lorentzian function \cite{griem1997} $L(\lambda,w_L,\lambda_{0,L})$ of half-width at half-maximum $\left(w_{\lambda,s,e,v}^{e',v'}\right)_{\text{col}}=(\lambda_{s,e,v}^{e',v'})^2\cdot(Z_{\text{opt},s,e,v}+Z_{\text{opt},s,e',v'})/(2\pi c)=:w_L$, centred at $\lambda=\lambda_{s,e,v}^{e',v'}=:\lambda_{0,L}$, where $Z_{\text{opt},s,e,v}$ is the optical collisional frequency per $s$-th species particle in the $(e,\,v)$ vibronic level with the other particles, and $Z_{\text{opt},s,e',v'}$ is the one with respect to the $(e',\,v')$ vibronic level. The former quantity is given by
\begin{equation}
Z_{\text{opt},s,e,v}=\left(\sum_{q\in\{\text{h}\},e'',v''}\frac{n_{q,e'',v''}\cdot\sigma_{\text{opt},s,e,v}^{q,e'',v''}}{1+\delta_{s,e,v}^{q,e'',v''}}\sqrt{\frac{8k_B T_{\text{tr}_\text{h}}}{\pi\,\mu_{s,q}}}\right)+n_\text{e}\cdot\sigma_{\text{opt},s,e,v}^{\text{e}}\sqrt{\frac{8k_B T_{\text{tr}_\text{e}}}{\pi\,m_{\text{e}}}}\text{ .}
\label{eq:Z_opt_s}
\end{equation}
The sum in \eqref{eq:Z_opt_s}  is done in all the heavy species,  their electronic levels and vibrational levels. The quantity $\delta_{s,e,v}^{q,e'',v''}$ is a Kronecker delta giving $1$ if $(s,e,v)=(q,e'',v'')$, and $0$ if not. The quantity $\mu_{s,q}=m_s\cdot m_q/(m_s+m_q)$ is the reduced mass of the collision partners. The term $\sigma_{\text{opt},s,e,v}^{q,e'',v''}$ corresponds to the optical collisional cross section for a collision between a particle of the $s$-th species in $(e,\,v)$ with a particle of the $q$-th species in $(e'',\,v'')$. And $\sigma_{\text{opt},s,e,v}^{\text{e}}$ corresponds to $\sigma_{\text{opt},s,e,v}^{q,e'',v''}$ for the case of the $q$-th species being a free electron. Penner \cite{penner1959} refers that, according to the available data, such quantities have the same order of magnitude as the respective collisional cross sections $\sigma_{s,e,v}^{q,e'',v''}$, and therefore, it was decided to consider $\sigma_{\text{opt},s,e,v}^{q,e'',v''}\approx\sigma_{s,e,v}^{q,e'',v''}$. No data were found for the case of collisions with free electrons, and, therefore, its contribution was disregarded. 

For the case of Stark broadening, it was decided to consider in this work the approach of Johnston \cite{johnston2006}, which solely accounts for the contribution of free electrons. The associated wavelength-specific line-shape factor $\left(\phi_{\lambda,s,e,v}^{e'v'}\right)_{\text{S}}$ is given by a Lorentzian function of half-width at half-maximum $\left(w_{\lambda,s,e,v}^{e',v'}\right)_{\text{S}}$, centred at $\lambda=\lambda_{s,e,v}^{e',v'}$.
The half-width at half-maximum regarded by Johnston has a form identical to the one of Park \cite{park1982}:
\begin{equation}
\left(w_{\lambda,s,e,v}^{e',v'}\right)_{\text{S}}=\left(w_{\lambda,s,e,v,\text{ref}}^{e',v'}\right)_{\text{S}}\left(\frac{T_{\text{tr}_\text{e}}}{T_{\text{tr}_\text{e},\text{ref}}}\right)^n\cdot\left(\frac{n_\text{e}}{n_{\text{e},\text{ref}}}\right)\text{ ,}
\label{eq:w_ll_Stark}
\end{equation}
where $\left(w_{\lambda,s,e,v,\text{ref}}^{e',v'}\right)_{\text{S}}$ is the half-width at half-maximum at a free electron translational temperature $T_{\text{tr}_\text{e}}=10,000\,\text{K}:=T_{\text{tr}_\text{e},\text{ref}}$ and at a free electron number density $n_\text{e}=10^{16}\,\text{cm}^{-3}:=n_{\text{e},\text{ref}}$. Park \cite{park1982} solely treated the case of argon, obtaining the exponent $n=0.33$ by fitting the curve \eqref{eq:w_ll_Stark} to the values of half-width at half-maximum $\left(w_{\lambda,s,e,v}^{e',v'}\right)_{\text{S}}$ at $n_\text{e}=n_{\text{e},\text{ref}}$ for different $T_{\text{tr}_\text{e}}$, issued by Griem \cite{griem1962,griem1964,griem1974}. Johnston considers this exponent value to be also acceptable for the case of the nitrogen and oxygen atoms. In the present work, it was decided to take such assumption one step further by regarding this same value for all the considered heavy species (\ch{N}, \ch{N+}, \ch{N2} and \ch{N2+}). Johnston then considers a model for the reference half-width at half-maximum of the form 
\begin{equation}
\left(w_{\lambda,s,e,v,\text{ref}}^{e',v'}\right)_{\text{S}}=\frac{C\cdot\left(\lambda_{s,e,v}^{e',v'}\right)^2}{\left(\epsilon_s^+-\epsilon_{s,\text{el-vib},e,v}\right)^n}\text{ ,}
\label{eq:w_ll_ref_Stark}
\end{equation}
with $\epsilon_s^+$ being the ionisation energy of the a $s$-th species particle from its ground level, and $C$ and $n$ some constants. The values for the ionisation energies of all the heavy species considered in this work were taken from the literature, being listed in Table \ref{tab:ionisation_energies}.
\begin{table}[H]
\setlength\tabcolsep{25pt} 
\centering
\caption{Ionisation energies from the ground level of \ch{N}, \ch{N+}, \ch{N2} and \ch{N2+}.}
\begin{scriptsize}
\begin{tabular}{ccc}
\toprule
$s$ & $\epsilon_s^+$ [eV] & Reference\\
\midrule
\ch{N} & 14.534 & Biémont \textit{et al.} \cite{biemont1999}\\
\ch{N+} & 29.601 & Biémont \textit{et al.} \cite{biemont1999}\\
\ch{N2} & 15.581 & Trickl \textit{et al.} \cite{trickl1989}\\
\ch{N2+} & 27.9 & Bahati \textit{et al.} \cite{bahati2001}\\ 
\bottomrule
\end{tabular}
\end{scriptsize}
\label{tab:ionisation_energies}
\end{table}
\noindent
Model \eqref{eq:w_ll_ref_Stark} is a variant of the one considered by Cowley in its theoretical work \cite{cowley1971}. Johnston obtained $C=1.69\times10^{10}\,[(\text{cm}^{-1})^{2.623}/\text{cm}^2]\cdot\text{cm}$ (here the ambiguous unit of energy $\text{cm}^{-1}$ was used which actually means $\text{cm}^{-1}\cdot h\cdot c$) and $n=2.623$ by fitting the curve \eqref{eq:w_ll_ref_Stark} to the values of $\left(w_{\lambda,s,e,v,\text{ref}}^{e',v'}\right)_{\text{S}}$ for the nitrogen and oxygen atoms, issued by Griem \cite{griem1974} and by Wilson and Nicolet \cite{wilson1967}.
%

For the case of resonance broadening, the theory developed by Griem \cite{griem1964} was considered. The wavelength-specific line-shape factor corresponds to a Lorentzian function centred at $\lambda=\lambda_{s,e,v}^{e',v'}$ and of half-width at half-maximum 
\begin{equation}
\left(w_{\lambda,s,e,v}^{e',v'}\right)_{\text{res}}=\frac{3}{32}\sqrt{\frac{g_{s,e,v}}{g_{s,e',v'}}}\frac{\left(\lambda_{s,e,v}^{e',v'}\right)^5\,A_{s,e,v}^{e',v'}}{\pi^3 c}n_{s,e',v'}\text{ ,}
\label{eq:w_nn_res}
\end{equation}
with $g_{s,e,v}$ being the degree of degeneracy of the $(e,\,v)$ vibronic level of the $s$-th species.

By noting that the convolution of two Lorentzian functions is also a Lorentzian function whose half-width at half-maximum and centre is the sum of them, one may show that the global line-shape factor defined by \eqref{eq:global_line_shape} corresponds to a Voigt function \cite{griem1997} $V(\lambda,w_G,w_L,\lambda_{0V})$ of Gaussian and Lorentzian half-widths at half-maxima $\left(w_{\lambda,s,e,v}^{e',v'}\right)_{\text{D}}=:w_G$ and $\left(w_{\lambda,s,e,v}^{e',v'}\right)_{\text{col}}+\left(w_{\lambda,s,e,v}^{e',v'}\right)_{\text{S}}+\left(w_{\lambda,s,e,v}^{e',v'}\right)_{\text{res}}=:w_L$, centred at $\lambda=\lambda_{s,e,v}^{e',v'}=:\lambda_{0,V}$.
The definition of the Voigt function involves an improper integral which is not analytically solvable, and therefore, one should consider a numerical method such as the trapezoidal integral rule, or an empirical approximation such as the formula proposed by Whiting \cite{whiting1968}, to account for it. Although the former may give better results, it requires much more computational resources than the latter. The empirical formula of  Whiting was then chosen, which according to his words, matches the exact function within 5 per cent at worst. The formula of Olivero and Longbothum \cite{olivero1977} for the half-width at half-maximum of the Voigt function $w_V$ was considered. It has an accuracy of about 0.01 per cent.

\section{Results}
\label{sec:results}

\subsection{The test matrix}
\label{subsection:test_matrix}

Brandis and Cruden \cite{brandis2018shock} issue data for a total of 42 shock tube  shots from which 17 are of benchmark quality (in some or all wavelength intervals) and may be used for validation of the developed numerical model. For the sake of compactness, it was decided to regard solely 3 of these 17 shots, more precisely, shots that spawned conditions of low, medium, and high speed hypersonic flows, and of benchmark quality in the four wavelength intervals (VUV, ``Blue'', ``Red'', and IR). These would allow a study about the dependence of the results on the upstream flow speed $u_{\infty}$ in the whole range of wavelength values. Under such criteria, the shots 40 (with $u_{\infty}=6.88\text{ m}/\text{s}$), 19 (with $u_{\infty}=10.32\text{ m}/\text{s}$), and 20 (with $u_{\infty}=11.16\text{ m}/\text{s}$) were taken.

\subsection{The analysis methodology}
\label{subsection:analysis_methodology}

In the following sections, the numerical and experimental results are compared, and possible causes for their discrepancies are enunciated. The dependence of the results on different adjustable parameters of the simulations is reported. To properly guide the reader over the extensive set of obtained results, it was decided to treat the three shots simultaneously for each of the four wavelength intervals and to show side-by-side the respective graphs for the two quantities that were measured in the experiments: the instrumentally resolved radiative intensities $\hat{I}^l$ and non-equilibrium metrics $\hat{I}_{\lambda}^{\text{ne},l}$, with $l\in\{\text{VUV},\text{``Blue''},\text{``Red''},\text{IR}\}$. Further comments on the behaviour of important physical quantities such as the translational temperatures and the species' mole fractions, as well as on the evolution of the system to equilibrium are provided. 

\subsection{Simulations of post-shock flows generated by a shock tube}
\label{subsection:0D_shock_tube_results}

\subsubsection{The case of the VUV radiation}
\label{subsubsection:1D_test_VUV}

\begin{figure}[H]
\centering
\centerline{\includegraphics[scale=1]{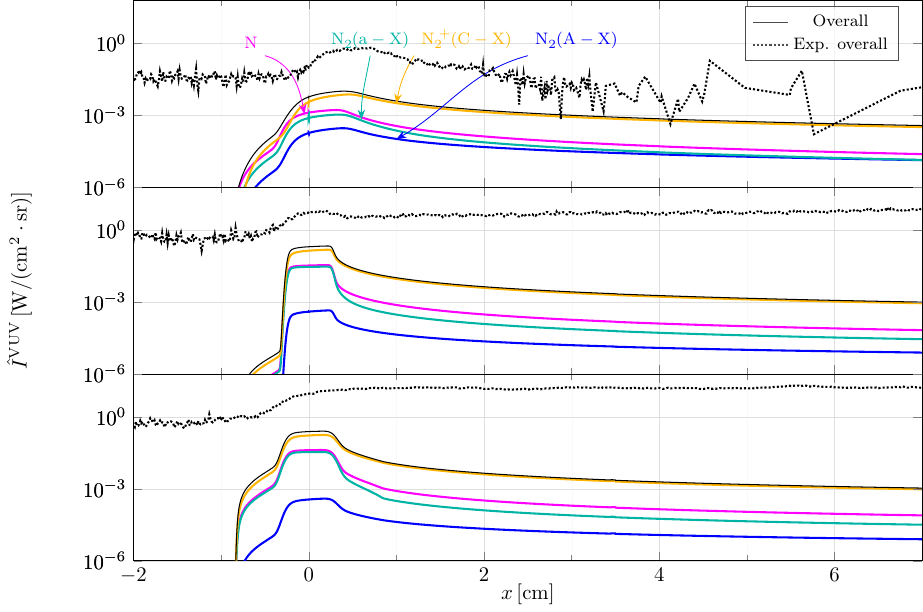}}
\vspace{-15pt}
\caption{Numerical (solid lines) and experimental (dotted lines) instrumentally resolved radiative intensities $\hat{I}^{\text{VUV}}(x)$ obtained for the low, medium and high speed shots.}
\label{fig:Results_I_inst_VUV_log}
\end{figure}

\begin{figure}[H]
\centering
\centerline{\includegraphics[scale=1]{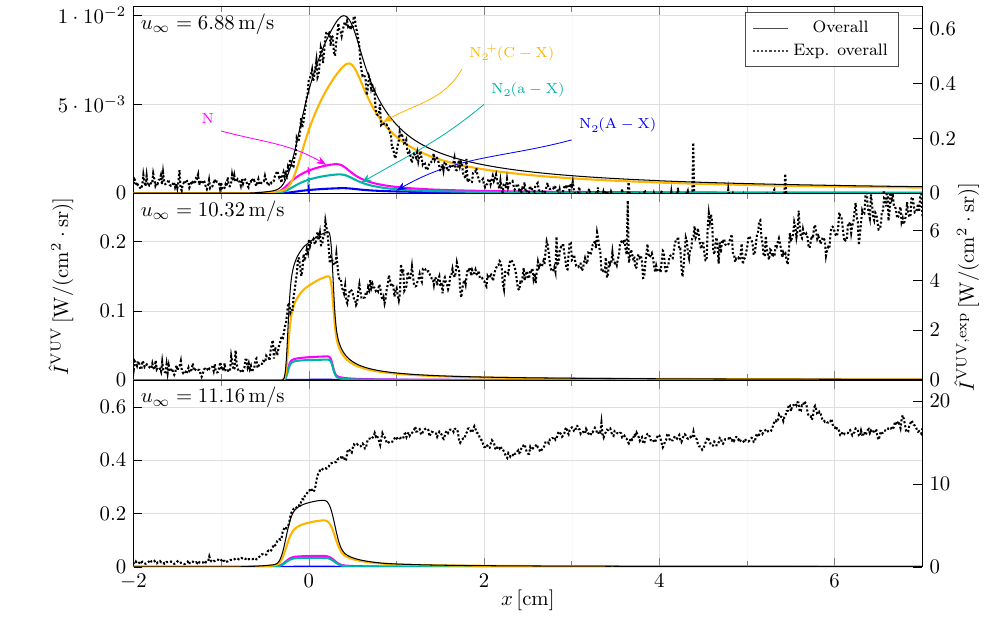}}
\vspace{-15pt}
\caption{Same as \Cref{fig:Results_I_inst_VUV_log}, but with two differently scaled $y$-axes: one for the numerical values (at left), and another for the experimental values (at right).}
\label{fig:Results_I_inst_VUV}
\end{figure}

\begin{figure}[H]
\centering
\centerline{\includegraphics[scale=1]{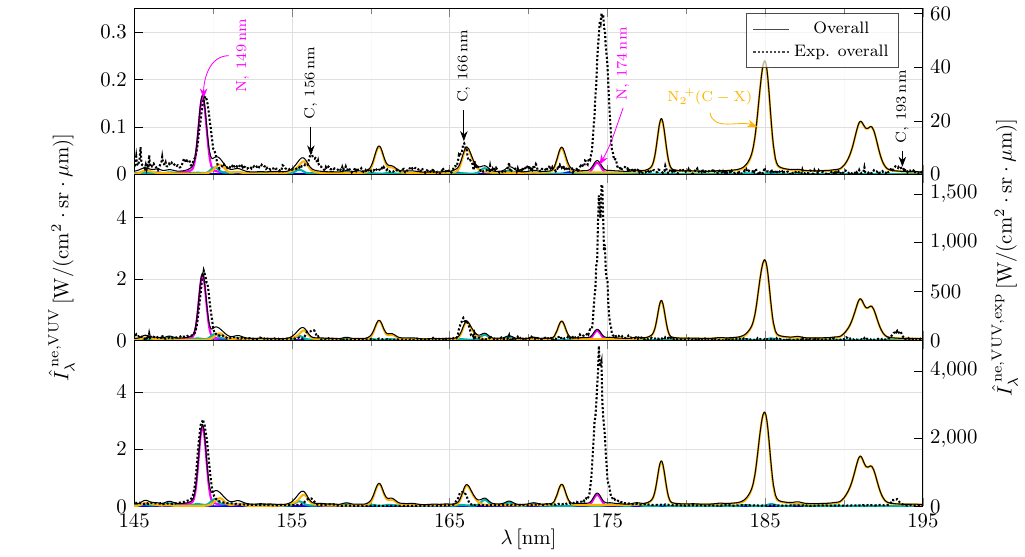}}
\vspace{-15pt}
\caption{Numerical (solid lines) and experimental (dotted lines) instrumentally resolved non-equilibrium metrics $\hat{I}_{\lambda}^{\,\text{ne},\text{VUV}}(\lambda)$ obtained for the low, medium and high speed shots.}
\label{fig:Results_I_ll_ne_inst_VUV}
\end{figure}

The instrumentally resolved radiative intensities $\hat{I}^{\text{VUV}}(x)$  obtained from the numerical simulations are depicted in \Cref{fig:Results_I_inst_VUV_log,fig:Results_I_inst_VUV}, and the non-equilibrium metrics $\hat{I}_{\lambda}^{\text{ne},\text{VUV}}(\lambda)$ are depicted in \Cref{fig:Results_I_ll_ne_inst_VUV}. The three graphs appearing in each figure are ordered by increasing free stream speed $u_{\infty}$ from top to bottom - the top one is with respect to shot $40$, the middle one to $19$, and the bottom one to $20$. The solid coloured lines represent the numerically obtained contributions of the different systems of spontaneous emission processes to $\hat{I}^{\text{VUV}}$ and $\hat{I}_{\lambda}^{\text{ne},\text{VUV}}$. The solid black lines represent the numerically obtained overall quantities (i.e. the sums of the contributions). The dotted black lines represent the experimentally obtained overall quantities. In \Cref{fig:Results_I_inst_VUV,fig:Results_I_ll_ne_inst_VUV} the numerical values are quantified in the left $y$-axis and the experimental ones in the right $y$-axis. These $y$-axes are scaled differently to make the two sets of values visually comparable. In the case of $\hat{I}^{\text{VUV}}$, the scales are such that the heights of the peaks match each other. And in the case of $\hat{I}_{\lambda}^{\text{ne},\text{VUV}}$, the scales are such that there is a coincidence between the peaks associated with spontaneous emission of \ch{N} at $\lambda=149\,\text{nm}$. The necessity of using different scales unveils immediately a significant discrepancy: the numerically obtained values are much lower than the experimental ones. In fact, the heights of the experimental $\hat{I}^{\text{VUV}}$ and $\hat{I}_{\lambda}^{\text{ne},\text{VUV}}$ peaks are one to two orders of magnitude higher than the numerical ones  (see \Cref{fig:Results_I_inst_VUV_log} for the case of the instumental radiative intensities). 

One should then analyse relative discrepancies, i.e. the discrepancies between the numerical and experimental results as if they were scaled to the same order of magnitude. Figure \ref{fig:Results_I_inst_VUV} reveals that the shape of the profile of $\hat{I}^{\text{VUV}}$ was well predicted for the case of the low speed shot. It also shows that spontaneous emissions of the type $\ch{N2+}(\text{C}-\text{X})$ were the main contribution to the numerical $\hat{I}^{\text{VUV}}$ values. However, the same cannot be said for the experimental counterpart: Figure \ref{fig:Results_I_ll_ne_inst_VUV} shows many $\hat{I}_{\lambda}^{\text{ne},\text{VUV}}$ peaks of considerable heigh assigned to $\ch{N2+}(\text{C}-\text{X})$ which were not actually observed in the experiment. Although the two peaks that appear at $\lambda=156\,\text{nm}$ and $\lambda=166\,\text{nm}$ seem to overlap others of $\ch{N2+}(\text{C}-\text{X})$, this should be mere coincidence. The two peaks may be predicted if one considers atomic carbon \ch{C} - a contaminant species - in the database, as Cruden and Brandis \cite{cruden2019} did. Furthermore, according to Cruden and Brandis \cite{cruden2019}, the experimental peak at $\lambda=193\,\text{nm}$ should also be due to \ch{C}. One finds the peak at $\lambda=149\,\text{nm}$ to be higher than the peak at $\lambda=174\,\text{nm}$ in the numerical spectra, but lower in the experimental one. These two peaks are due to spontaneous emission of \ch{N}. Such discrepancy may be due to the absorption of radiation associated with the first peak being greater than the one associated with the second peak in the experiment. Meanwhile, in the simulation, absorption was completely disregarded. The discrepancy may also be due to an improper modelling of the rate coefficients which dictate the population of the nitrogen atoms that spontaneously emit radiation at wavelengths $\lambda=149\,\text{nm}$ and $\lambda=174\,\text{nm}$. One way of making the ratio between the heights of the two peaks more coherent with the experimental result is by decreasing the rate of excitation of atomic nitrogen to the upper level of the system associated with the first peak and increasing the one associated with the second. However, in the present case, the upper levels of the system associated with the two peaks are the same, corresponding to the fifth level, say\footnote{The electronic levels of the atomic particles will be labelled here by integer numbers $e=0,\,1,\,...$} $\ch{N}(4)$. The first peak is a result of the transitions $\ch{N}(4-1)$, and the second is a result of the transitions $\ch{N}(4-2)$. Therefore, changing the values of the rate coefficients for excitation would not make any difference to the ratio between the heights of the peaks. 
%

The experimental instrumentally resolved radiative intensity $\hat{I}^{\text{VUV}}$ obtained for the case of the medium speed shot has a peak proceeded by a non-zero plateau. This plateau was not predicted at all. A transition to zero occurs instead. And for the case of the high speed shot, the experiment did not produce a sole peak, but a coalescence between a peak and a plateau surpassing it, which was also not predicted.
%

\subsubsection{The case of the ``Blue'' radiation}
\label{subsubsection:1D_test_Blue}
\begin{figure}[H]
\centering
\centerline{\includegraphics[scale=1]{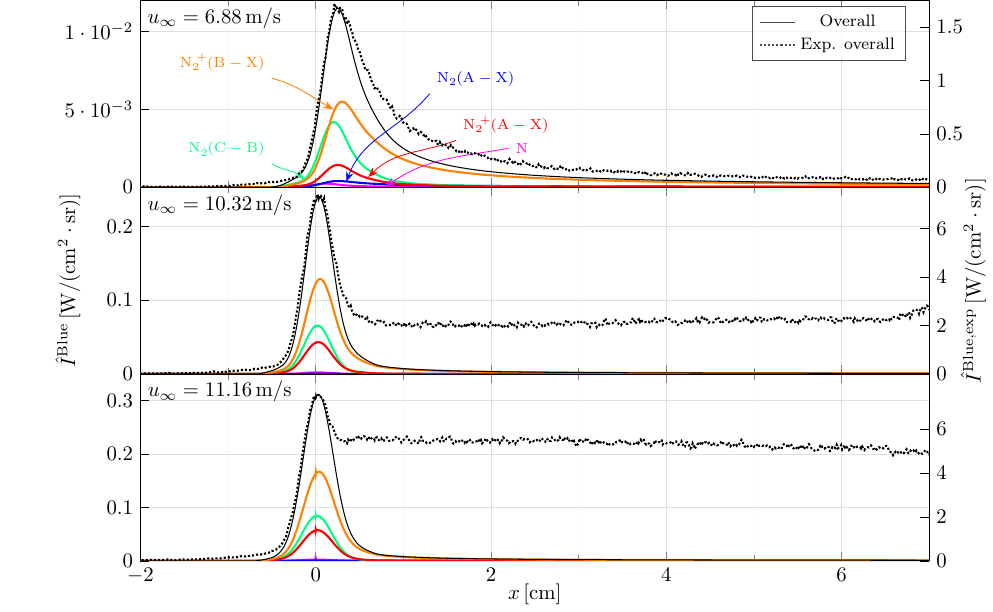}}
\vspace{-15pt}
\caption{Numerical (solid lines) and experimental (dotted lines) instrumentally resolved radiative intensities $\hat{I}^{\text{Blue}}(x)$ obtained for the low, medium and high speed shots.}
\label{fig:Results_I_inst_Blue}
\end{figure}

\begin{figure}[H]
\centering
\centerline{\includegraphics[scale=1]{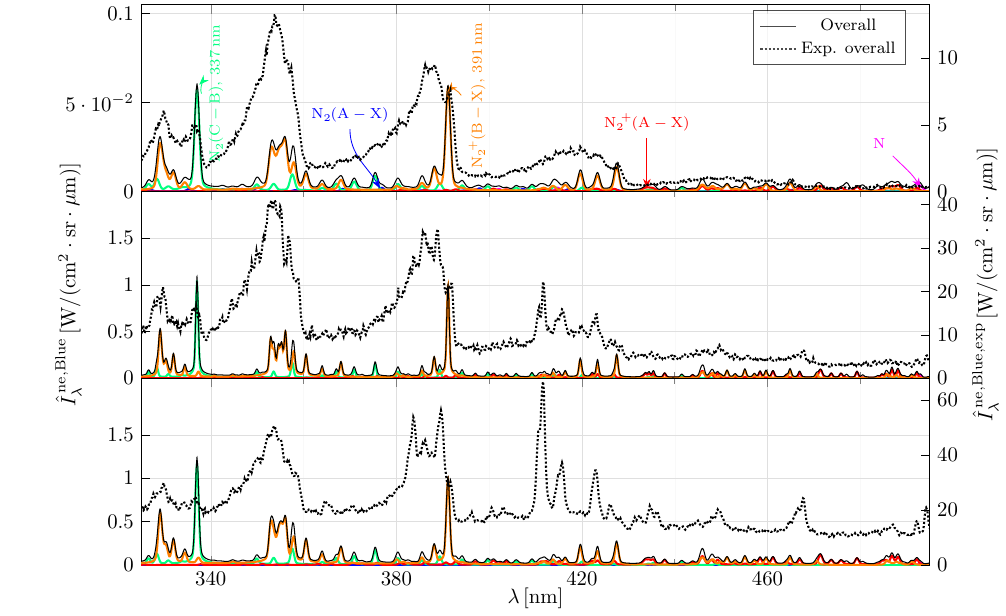}}
\vspace{-15pt}
\caption{Numerical (solid lines) and experimental (dotted lines) instrumentally resolved non-equilibrium metrics $\hat{I}_{\lambda}^{\,\text{ne},\text{Blue}}(\lambda)$ obtained for the low, medium and high speed shots.}
\label{fig:Results_I_ll_ne_inst_Blue}
\end{figure}

The obtained instrumentally resolved radiative intensities $\hat{I}^{\text{Blue}}(x)$ and non-equilibrium metrics $\hat{I}_{\lambda}^{\text{ne},\text{Blue}}(\lambda)$ are depicted in Figure 
\ref{fig:Results_I_inst_Blue} and Figure \ref{fig:Results_I_ll_ne_inst_Blue}, respectively. The main contributors to the overall quantities are $\ch{N2+}(\text{B}-\text{X})$, $\ch{N2}(\text{C}-\text{B})$ and $\ch{N2+}(\text{A}-\text{X})$, and the residual contributors correspond to \ch{N} and $\ch{N2}(\text{A}-\text{X})$. Meanwhile, Cruden and Brandis \cite{cruden2019} showed that $\ch{N2}(\text{C}-\text{B})$, $\ch{N2+}(\text{B}-\text{X})$, \ch{N}, and the contaminant species \ch{CN} (cyanogen radical) are enough to describe the experimentally obtained spectrum for the case of the medium speed shot. 

As for the case of the VUV radiation, the instrumentally resolved radiative intensities and non-equilibrium metrics were underestimated by one to two orders of magnitude. The scales of the $y$-axes of the non-equilibrium metrics graphs were chosen such that there was a coincidence between the numerical and experimental peaks associated with spontaneous emission of $\ch{N2+}(\text{B}-\text{X})$ at $\lambda=391\,\text{nm}$. The profile of the numerical radiative intensity $\hat{I}^{\text{Blue}}$ obtained in the low speed shot corresponds to a peak in which the transition from the maximum to zero is faster than in the experimental counterpart. For the case of the medium and high speed shots, the shapes of the peaks were remarkably well predicted, contrary to the plateaus that proceed them: the numerical radiative intensities transit to zero instead of converging to these plateaus. Such behaviour was also obtained in the case of the VUV radiation. 

Sets of peaks appearing in the experimental non-equilibrium metrics $\hat{I}_{\lambda}^{\text{ne},\text{Blue}}$ may be discerned. These are in some way predicted by the numerical model. However, the heights of the individual peaks for each set do not agree particularly well with the experimental ones, namely, the ones associated with spontaneous emissions of the types $\ch{N2}(\text{C},0-\text{B},0)$ and $\ch{N2+}(\text{B},0-\text{X},0)$, at $\lambda=337\,\text{nm}$ and $\lambda=391\,\text{nm}$, respectively. These were undoubtedly overestimated, prevailing over the others. The discrepancies between the numerical and experimental non-equilibrium metrics may in some part be due to the regarded spontaneous emissions being vibronic-specific instead of rovibronic. If rovibronic-specific spontaneous emissions were considered, the peaks associated with the vibronic levels would be divided into several others (each one associated with a rotational level) whose centres may be more or less far from each other, depending on the energy of the rotational levels. The importance of the rotational levels to the spectra should be accessed in the future.

\subsubsection{The case of the ``Red'' radiation}
\label{subsubsection:1D_test_Red}
\begin{figure}[H]
\centering
\centerline{\includegraphics[scale=1]{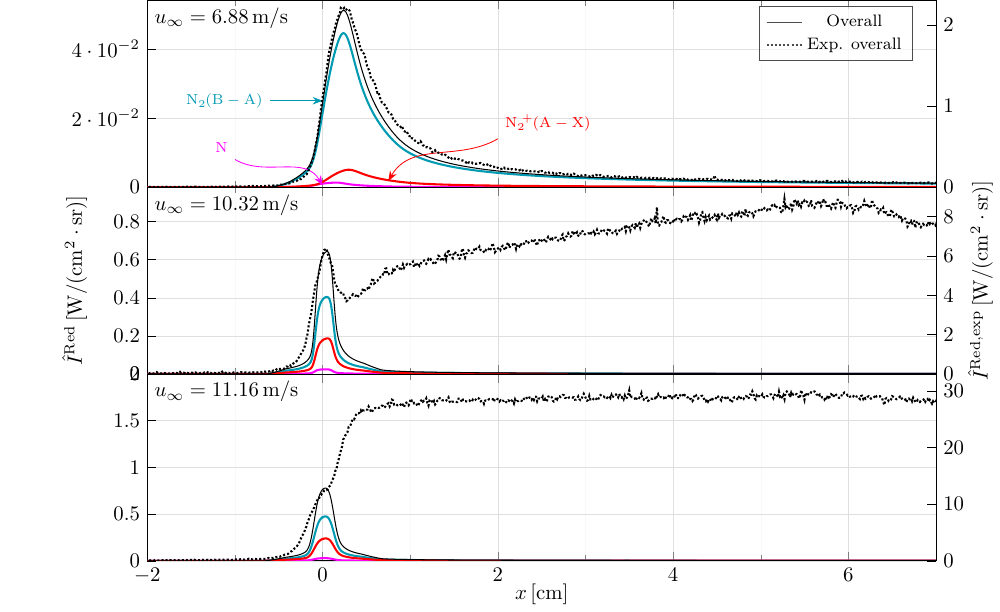}}
\vspace{-15pt}
\caption{Numerical (solid lines) and experimental (dotted lines) instrumentally resolved radiative intensities $\hat{I}^{\text{Red}}(x)$ obtained for the low, medium and high speed shots.}
\label{fig:Results_I_inst_Red}
\end{figure}

\begin{figure}[H]
\centering
\centerline{\includegraphics[scale=1]{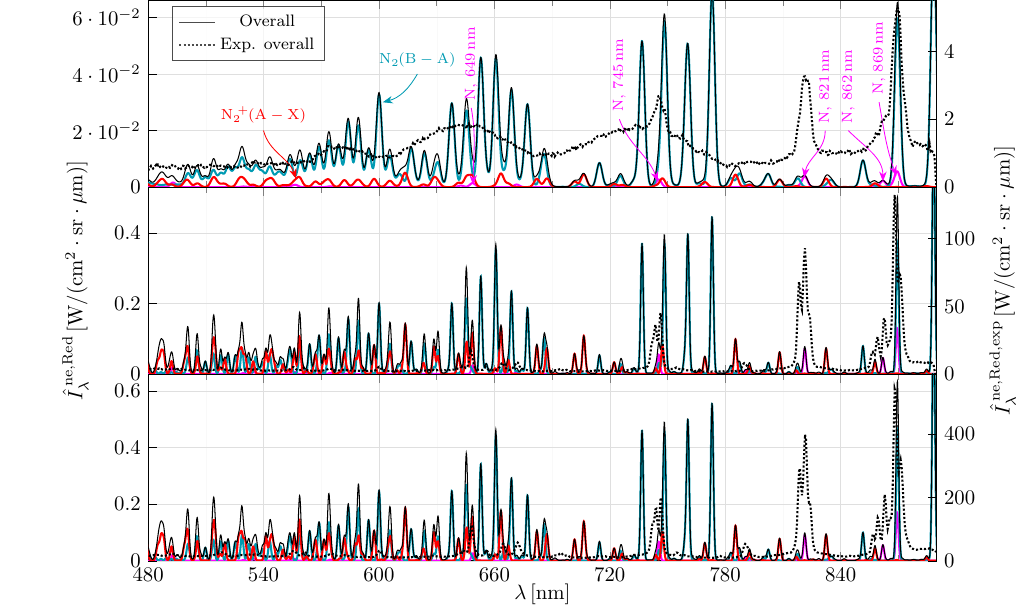}}
\vspace{-15pt}
\caption{Numerical (solid lines) and experimental (dotted lines) instrumentally resolved non-equilibrium metrics $\hat{I}_{\lambda}^{\,\text{ne},\text{Red}}(\lambda)$ obtained for the low, medium and high speed shots.}
\label{fig:Results_I_ll_ne_inst_Red}
\end{figure}

Figures \ref{fig:Results_I_inst_Red} and \ref{fig:Results_I_ll_ne_inst_Red} depict the obtained instrumentally resolved radiative intensities $\hat{I}^{\text{Red}}(x)$ and non-equilibrium metrics $\hat{I}_{\lambda}^{\text{ne},\text{Red}}(\lambda)$, respectively. The scales of the $\hat{I}_{\lambda}^{\text{ne},\text{Red}}$ graphs are such that the experimental and numerical peaks at $\lambda=869\,\text{nm}$ match each other. The heights of the experimental $\hat{I}^{\text{Red}}$ and $\hat{I}_{\lambda}^{\text{ne},\text{Red}}$ peaks were underestimated by one to two orders of magnitude. As happened for the case of the ``Blue'' radiation, in the low speed shot, the transition of the numerical radiative intensity $\hat{I}_{\lambda}^{\text{ne},\text{Red}}$ from the peak maximum value to zero was slighter faster than the experimental one. For the medium and high speed shots, plateaus proceeding peaks were observed experimentally. These were not predicted by the numerical model. 

There is strong evidence of a relative underestimation of spontaneous emissions of \ch{N}, since the most prominent peaks (such as the ones at $\lambda=649$, $745$, $821$, $862$ and $869\,\text{nm}$) from the experimental spectra, which are due to \ch{N}, are exceeded in the numerical spectra by the ones associated with $\ch{N2}(\text{B}-\text{A})$. Also, spontaneous emissions of the type $\ch{N2+}(\text{A}-\text{X})$ should not be as relevant as they were found to be in the obtained numerical spectra. For the case of the medium speed shot, Cruden and Brandis \cite{cruden2019} solely considered spontaneous emissions from \ch{N} (the dominant ones), $\ch{N2}(\text{B}-\text{A})$ and \ch{H}(as a contaminant species) in this wavelength interval.


\subsubsection{The case of the IR radiation}
\label{subsubsection:1D_test_IR}
\begin{figure}[H]
\centering
\centerline{\includegraphics[scale=1]{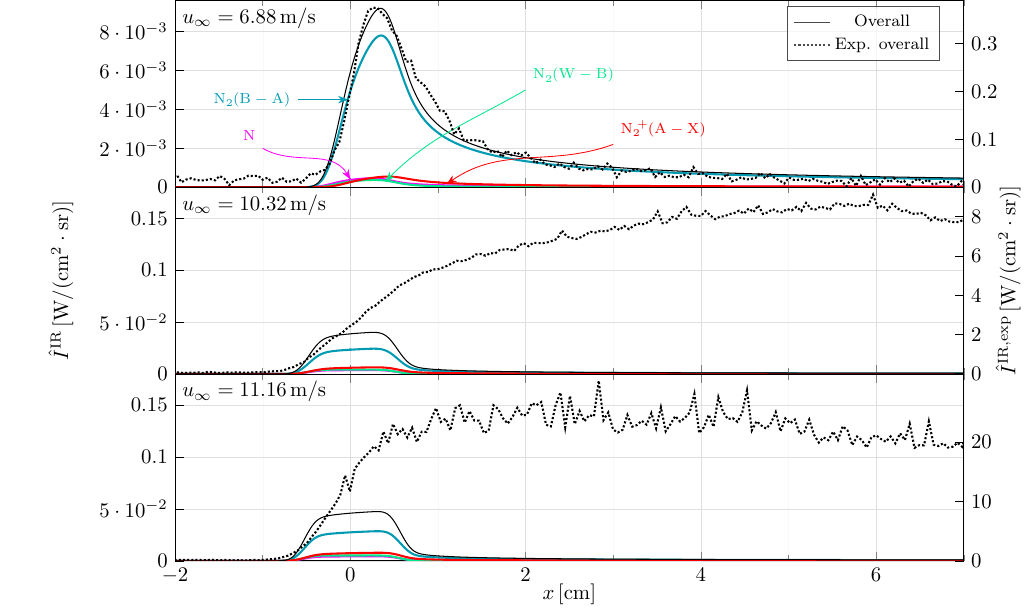}}
\vspace{-15pt}
\caption{Numerical (solid lines) and experimental (dotted lines) instrumentally resolved radiative intensities $\hat{I}^{\text{IR}}(x)$ obtained for the low, medium and high speed shots.}
\label{fig:Results_I_inst_IR}
\end{figure}

\begin{figure}[H]
\centering
\centerline{\includegraphics[scale=1]{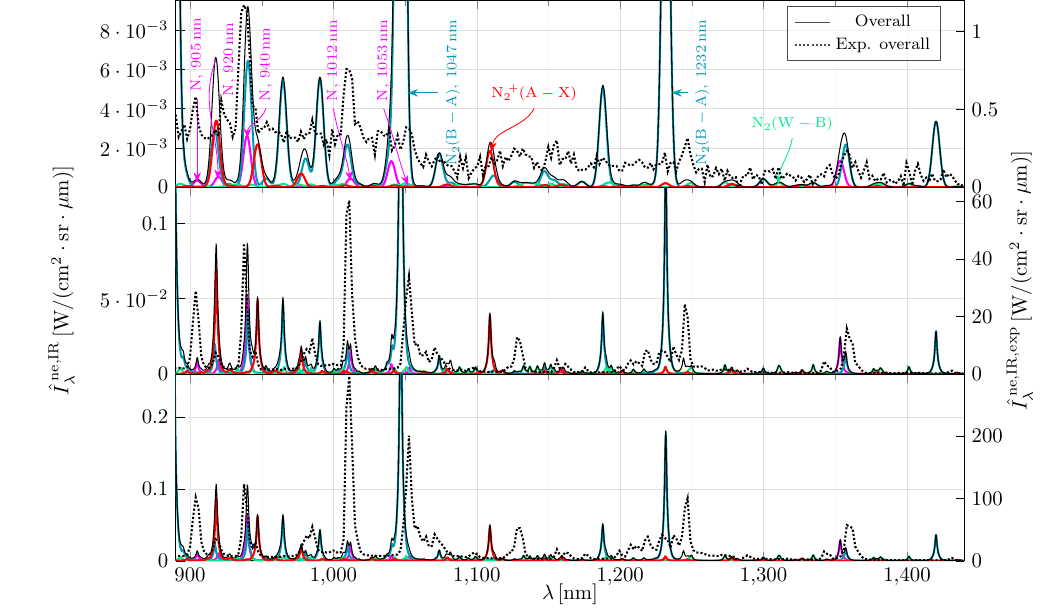}}
\vspace{-15pt}
\caption{Numerical (solid lines) and experimental (dotted lines) instrumentally resolved non-equilibrium metrics $\hat{I}_{\lambda}^{\,\text{ne},\text{IR}}(\lambda)$ obtained for the low, medium and high speed shots.}
\label{fig:Results_I_ll_ne_inst_IR}
\end{figure}

Figures \ref{fig:Results_I_inst_IR} and \ref{fig:Results_I_ll_ne_inst_IR} depict the obtained instrumentally resolved radiative intensities $\hat{I}^{\text{IR}}(x)$ and non-equilibrium metrics $\hat{I}_{\lambda}^{\text{ne},\text{IR}}(\lambda)$, respectively. The scales of the $\hat{I}_{\lambda}^{\text{ne},\text{IR}}$ graphs are such that the experimental and numerical peaks at $\lambda=940\,\text{nm}$ match each other. As for the other wavelength intervals, it was found that the numerical model underestimated the radiative intensities and non-equilibrium metrics by one to two orders of magnitude. In the case of the low speed shot, the shape of the profile of $\hat{I}^{\text{IR}}$ was well predicted. For the cases of medium and high speed shots, individual peaks cannot be discerned from the experiment but coalescences between peaks and plateaus occurring above them. Again, such phenomena was not predicted by the numerical model.

The obtained non-equilibrium metrics show that spontaneous emissions of the type $\ch{N2}(\text{B}-\text{A})$ seem to be relatively overestimated, specially the ones at $\lambda=1047\,\text{nm}$ - due to transitions from $v=0$ to $v'=0$ - and $\lambda=1232\,\text{nm}$ - due to transitions from $v=0$ to $v'=1$ - which prevail over the others. Conversely, the peaks associated with spontaneous emission of \ch{N} (such as the ones at $\lambda=905\,\text{nm}$, $\lambda=940\,\text{nm}$, $\lambda=1012\,\text{nm}$ and $\lambda=1053\,\text{nm}$) seem to be relatively underestimated, since these predominate in the experimental spectra but not in the numerical spectra.

\subsection{Mole fractions, temperatures, and evolution to equilibrium}
\label{subsection:results_mole_temperatures_equilibrium}

\begin{figure}[H]
\centering
\centerline{\includegraphics[scale=1]{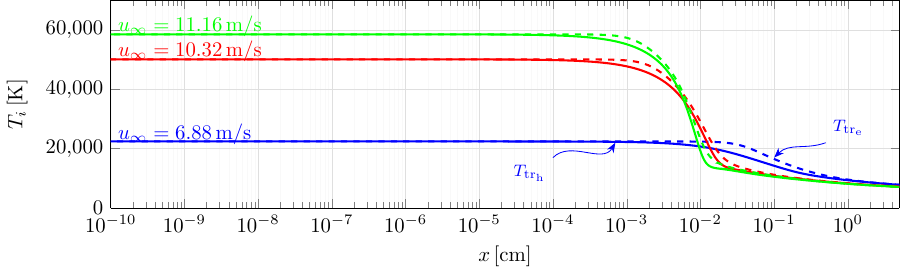}}
\vspace{-15pt}
\caption{Heavy particle (solid lines) and free electron (dashed lines) translational temperatures $T_{\text{tr}_\text{h}}(x)$ and $T_{\text{tr}_\text{e}}(x)$, obtained for the low (blue), medium (red) and high (green) speed shots.}
\label{fig:Results_T_1D}
\end{figure}

\begin{figure}[H]
\centering
\centerline{\includegraphics[scale=1]{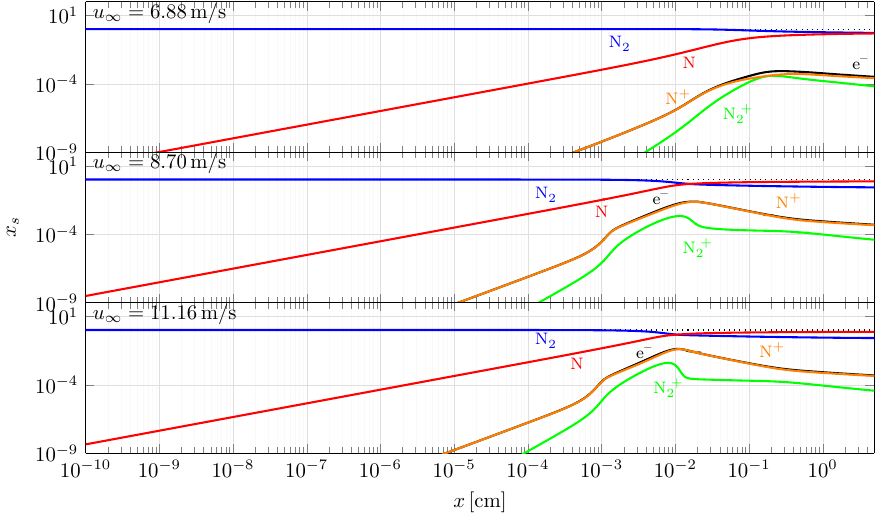}}
\vspace{-15pt}
\caption{Species-specific mole fractions $x_s(x)$ obtained for the low, medium and high speed shots.}
\label{fig:Results_x_s_1D}
\end{figure}

Figure \ref{fig:Results_T_1D} depicts the predicted heavy particle and free electron translational temperatures, $T_{\text{tr}_\text{h}}(x)$ and $T_{\text{tr}_\text{e}}(x)$. According to the Rankine-Hugoniot condition, the values of the translational temperatures immediately downstream of the shock wave ($x=0$) correspond to $T_{\text{tr}_\text{h},2}=T_{\text{tr}_\text{e},2}=22,434\,\text{K}$, $50,122\,\text{K}$ and $58,565\,\text{K}$ for the low, medium and high speed shots, respectively. As expected, these decrease with $x$, due to the endothermic processes, to somewhat identical values, and with a rate which is as high as the immediately downstream temperature. 
%
%
Figure \ref{fig:Results_x_s_1D} presents the evolutions of the mole fractions $x_s(x)$ of the five species considered in the simulations - \ch{N2}, \ch{N}, \ch{N2+}, \ch{N+} and \ch{e-} - for the low, medium and high speed shots. One finds that an higher upstream speed implies an earlier dissociation and ionisation of the particles. In the medium and high speed shots, the dissociation is such that the mole fraction of atomic nitrogen \ch{N} surpasses the mole fraction of molecular nitrogen \ch{N2}. There is an increase and posterior decrease of the mole fractions of atomic nitrogen ions \ch{N+}, molecular nitrogen ions \ch{N2+} and free electrons \ch{e-}. The former is associated with ionisation and the latter to recombination. The higher the upstream speed, the more abrupt the increases and decreases of the mole fractions of the charged particles are. 
%
%
Table \ref{tab:eq_conditions} makes a synopsis on the temperatures $T_{\text{tr}_\text{h}}$ and $T_{\text{tr}_\text{e}}$, and mole fractions $x_s$, obtained from the simulations of the three shots, at a downstream point far from the shock wave, $x=5\,\text{cm}$ - where the radiation intensities were found to be stagnated. The table also shows the values issued by Cruden and Brandis \cite{cruden2019} for the medium speed shot. These were obtained by fitting the conditions of a hypothetical system such that the computed ``Blue'' and ``Red'' spectra matched the experimental ones, using the NEQAIR tool \cite{NEQAIR96,cruden2014updates}, at a $x$ position for which the plateaus of radiative intensities occurred. With that objective, a two-temperature model - of temperatures $T_{\text{tr}_\text{h}-\text{rot}}$ and $T_{\text{vib}-\text{el}-\text{tr}_\text{e}}$ - was assumed and the populations of some particular energy levels were adjusted. Regarding the values of the temperatures at $x=5\,\text{cm}$ obtained in the present work, one can say that the heavy particle translational temperature departed from the free electron translational temperature by $-18\,\text{K}$ (or $-0.23\,\%$), $-3\,\text{K}$ (or $-0.04\,\%$) and $-1\,\text{K}$ (or $-0.03\,\%$) for the cases of the low, medium and high speed shots. These values are sufficiently small for one to assume that an equilibrium between the respective energy modes was attained. The values of the temperatures obtained by Cruden and Brandis are significantly higher than the ones of this work. The analysis of the ``Blue'' spectra resulted in a $T_{\text{tr}_\text{h}}$ value which is greater by $\left(1,986 \pm 170\right)\,\text{K}$ (or $\left(28\pm2\right)\,\%$) and a $T_{\text{tr}_\text{e}}$ value which is greater by $\left(3,213 \pm 300\right)\,\text{K}$ (or $\left(45\pm4\right)\,\%$). The analysis of the ``Red'' spectra resulted in a $T_{\text{tr}_\text{h}}$ value which is greater by $\left(2,606 \pm 2,490\right)\,\text{K}$ (or $\left(36\pm35\right)\,\%$) and a $T_{\text{tr}_\text{e}}$ value which is greater by $\left(2,013\pm 130\right)\,\text{K}$ (or $\left(28\pm2\right)\,\%$).
The mole fraction of \ch{N} obtained from the simulation of the medium speed shot in the present work, at $x=5\,\text{cm}$, is almost the double of the one obtained from the simulation of the low speed shot, showcasing a very meaningful increase on the dissociation of \ch{N2}. The mole fraction of \ch{N2+} is lower (by $38\,\%$) - due to a much faster recombination - and the one of \ch{N+} is almost the double. The differences between the values of the mole fractions obtained from the simulation of the high speed shot and the ones obtained from the simulation of the medium speed shot are not significant enough, as also are not the differences between the upstream speeds. Regarding the results of the work of Cruden and Brandis \cite{cruden2019} for the medium speed shot, one may say that not only the differences between the obtained chemical compositions in that work and the present one are staggering but also the differences between the chemical compositions resulting from the analysis of the ``Blue'' and ``Red'' spectra in the former. The results of the analysis of the ``Blue'' spectra showed a greater importance of the molecular particles \ch{N2} and \ch{N2+} and free electrons \ch{e-} in the composition whereas the results of the analysis of the ``Red'' spectra showed a greater importance of atomic nitrogen \ch{N}. In fact, the mole fractions of \ch{N2}, \ch{N2+} and \ch{e-} obtained in the the former case are $22$, $133$ and $30$ times greater than the ones obtained in the latter case, and the mole fraction of \ch{N} is $57\,\%$ lower. This incoherence evidences that the models regarded in NEQAIR may not represent sufficiently well the reality. The results of Cruden and Brandis also evidence a much stronger dissociation of \ch{N2} and ionisation of \ch{N} than the ones of the present work: the predicted mole fraction of \ch{N2} is at least $9$ times lower, and the mole fraction of \ch{N+} is at least $20$ times greater. Moreover, this greater dissociation and ionisation are accomplished with a much lower cost of heavy particle and free electron translational temperatures. Such result means that the post-shock elements of fluid simulated in the present work may have lost more energy than they should, or did not gain as much. Since it was found that the radiation variables were underpredicted by several orders of magnitude, the former hypothesis should be disregarded. It is then suspected that some source of energy which is external to the regarded system occurs in the experiment. Also, the unaccounted absorption of radiation and conduction of heat within the system may have some impact in the redistribution of the energy. 
\begin{table}[H]
\setlength\tabcolsep{3pt} 
\centering
\begin{threeparttable}
\caption{Temperatures $T_{\text{tr}_\text{h}}$ and $T_{\text{tr}_\text{e}}$, and mole fractions $x_s$, at $x=5\,\text{cm}$, obtained from the simulations of the low, medium and high speed shots, as well as the ones obtained by Cruden and Brandis \cite{cruden2019}.}
\begin{scriptsize}
\centerline{
\begin{tabular}{c|cccccccccc}
\toprule
$u_\infty[\text{m}/\text{s}]$ & Ref. & $T_{\text{tr}_\text{h}}[\text{K}]$ & $T_{\text{tr}_\text{e}}[\text{K}]$ & $x_{\ch{N2}}$ & $x_{\ch{N}}$ & $x_{\ch{N2+}}$ & $x_{\ch{N+}}$ & $x_{\ch{e-}}$ & $x_{\ch{CN}}$ & $x_{\ch{H}}$\\
\midrule
$6.88$ & This work & $7,793$ & $7,811$ & $0.52$ & $0.48$ & $6.5(-5)$ & $2.6(-4)$ & $3.2(-4)$ & --- & ---\\
\greymidrule
\multirow{3}{*}{$10.32$} & This work & $7,184$ & $7,187$ & $0.27$ & $0.73$ & $4.0(-5)$ & $4.5(-4)$ & $4.9(-4)$ & --- & ---\\
 & NEQAIR, ``Blue'' \cite{cruden2019} & $9,170\pm 170$ & $10,400\pm 300$ & $2.9(-2)$ & $0.42$ & $1.6(-2)$ & $0.27$ & $0.27$ & $7.3(-4)$ & ---\\
 & NEQAIR, ``Red'' \cite{cruden2019} & $9,790\pm 2,490$ & $9,200\pm 130$ & $1.3(-3)$ & $0.98$ & $1.2(-4)$ & $8.9(-3)$ & $8.9(-3)$ & --- & $1.0(-3)$\\
\greymidrule
$11.16$ & This work & $7,098$ & $7,100$ & $0.26$ & $0.74$ & $3.7(-5)$ & $4.4(-4)$ & $4.7(-4)$ & --- & ---\\ 
\bottomrule
\end{tabular}
}
\end{scriptsize}
\label{tab:eq_conditions}
\begin{scriptsize}
\begin{tablenotes}
\item{The number between parenthesis in the mole fractions cells correspond to the orders of magnitude of these quantities.}
\end{tablenotes}
\end{scriptsize}
\end{threeparttable}
\end{table}

To ascertain the impact of energy loss by radiation on the developed conditions of the post-shock flow, it was decided to perform simulations disregarding spontaneous emission processes and comparing the respective results with the ones including these processes. Table \ref{tab:eq_conditions_no_se} makes a synopsis on the resultant temperatures $T_{\text{tr}_\text{h}}$ and $T_{\text{tr}_\text{e}}$, and mole fractions $x_s$, at $x=5\,\text{cm}$. All of the temperatures are higher than the respective ones obtained when regarding spontaneous emission processes, and increasing with the upstream speed. While the differences are negligible for the case of the low speed shot - $T_{\text{tr}_\text{h}}$ differs by just $266\,\text{K}$ (or $3.4\,\%$) - they are meaningful for the case of the medium and high speed shots - $T_{\text{tr}_\text{h}}$ differs by $2080\,\text{K}$ (or $29\,\%$)  and $3,016\,\text{K}$ (or $42\,\%$). Meanwhile, a greater dissociation and ionisation have occurred, as evidenced by the lower value for the mole fraction of \ch{N2} and the higher values for the mole fractions of \ch{N} and \ch{N+}. These results show that the energy lost by spontaneous emission in the highly excited systems has a much higher proportion than in the less excited ones. The more excited the particles are, the higher the number of spontaneous emissions and the higher the energy that is lost through radiation. Surprisingly, the conditions for the medium speed shot disregarding spontaneous emission processes are now coherent with the ones obtained by Cruden and Brandis \cite{cruden2019}, with the exception of the mole fraction of \ch{N2+} which is substantially  lower. This result sheds light on the previously enunciated hypothesis of the simulated elements of fluid not receiving the energy that their counterparts seem to receive in the experiment.

\begin{table}[H]
\centering
\caption{Temperatures $T_{\text{tr}_\text{h}}$ and $T_{\text{tr}_\text{e}}$, and mole fractions $x_s$, obtained from the simulations of the low, medium and high speed shots, at $x=5\,\text{cm}$, disregarding spontaneous emission processes.}
\begin{scriptsize}
\begin{tabular}{c|ccccccc}
\toprule
$u_\infty[\text{m}/\text{s}]$ & $T_{\text{tr}_\text{h}}[\text{K}]$ & $T_{\text{tr}_\text{e}}[\text{K}]$ & $x_{\ch{N2}}$ & $x_{\ch{N}}$ & $x_{\ch{N2+}}$ & $x_{\ch{N+}}$ & $x_{\ch{e-}}$\\
\midrule
$6.88$ & $8,059$ & $8,076$ & $0.46$ & $0.54$ & $1.0\times10^{-4}$ & $4.6\times10^{-4}$ & $5.7\times10^{-4}$\\
$10.32$ & $9,264$ & $9,265$ & $0.022$ & $0.93$ & $3.1\times10^{-5}$ & $0.022$ & $0.022$\\
$11.16$ & $10,114$ & $10,114$ & $0.0041$ & $0.89$ & $2.4\times10^{-5}$ & $0.052$ & $0.052$ \\ 
\bottomrule
\end{tabular}
\end{scriptsize}
\label{tab:eq_conditions_no_se}
\end{table}

Another important result that may be ascertained from the numerical simulations is the systems departure from thermal equilibrium. For that purpose, representative temperatures were computed. The representative vibrational temperature associated with the $e$-th electronic level of the $s$-th species, $T_{s,\text{vib},e}$, was obtained by fitting the curve
\begin{equation}
\ln\left(\frac{x_{s,e,v}}{g_{s,\text{el},e}\cdot g_{s,\text{vib},e,v}}\right)=-\frac{1}{k_B T_{s,\text{vib},e}}\epsilon_{s,\text{vib},e,v}+\ln\left(\frac{x_{s,e}}{g_{s,\text{el},e} \cdot Q_{s,\text{vib},e}(T_{s,\text{vib},e})}\right)\text{ ,}
\label{eq:T_s_vib_rep_2}
\end{equation}
where $Q_{s,\text{vib},e}$ is the vibrational partition function associated with the $e$-th electronic level of the $s$-th species, to the points of abscissae $\epsilon_{s,\text{vib},e,v}$ and ordinates $\ln(x_{s,e,v}/(g_{s,\text{el},e}\cdot g_{s,\text{vib},e,v}))$. Equation \eqref{eq:T_s_vib_rep_2} tells that under vibrational self-equilibrium, the natural logarithm of the mole fraction of $s$-th species particles in a state of the $e$-th electronic level and $v$-th vibrational level, i.e. $\ln(x_{s,e,v}/(g_{s,\text{el},e}\cdot g_{s,\text{vib},e,v}))$, decreases linearly with their vibrational energy $\epsilon_{s,\text{vib},e,v}$. The representative electronic temperature associated with the $s$-th species, $T_{s,\text{el}}$, may be obtained by fitting the curve
\begin{equation}
\ln\left(\frac{x_{s,e}}{g_{s,\text{el},e}}\right)=-\frac{1}{k_B T_{s,\text{el}}}\epsilon_{s,\text{el},e}+\ln\left(\frac{x_{s}\cdot Q_{s,\text{vib},e}(T_{s,\text{vib},e})}{Q_{s,\text{el-vib}}(\{T_{s,\text{vib},e}\},T_{s,\text{el}})}\right)\text{ ,}
\label{eq:T_s_el_rep_2}
\end{equation}
where $Q_{s,\text{el-vib}}$ is the vibronic partition function associated with the $s$-th species, to the points of abscissae $\epsilon_{s,\text{el},e}$ and ordinates $\ln(x_{s,e}/g_{s,\text{el},e})$. In thermal equilibrium conditions, one has $T_{\text{tr}_\text{h}}=T_{\text{tr}_\text{e}}=T_{s,\text{vib},e}=T_{s,\text{el}}=:T$, $\forall$ $s$ and $e$. By inserting \eqref{eq:T_s_el_rep_2} into \eqref{eq:T_s_vib_rep_2} one gets in such conditions
\begin{equation}
\ln\left(\frac{x_{s,e,v}}{g_{s,\text{el},e}\cdot g_{s,\text{vib},e,v}}\right)=-\frac{1}{k_B T}\left(\epsilon_{s,\text{el},e}+\epsilon_{s,\text{vib},e,v}\right)+\ln\left(\frac{x_{s}}{Q_{s,\text{el-vib}}(T)}\right)\text{ .}
\label{eq:T_eq_cond}
\end{equation}
Therefore, at thermal equilibrium, the points of abscissae $\epsilon_{s,\text{el},e}+\epsilon_{s,\text{vib},e,v}=:\epsilon_{s,\text{el-vib},e,v}$ and ordinates $\ln(x_{s,e,v}/(g_{s,\text{el},e}\cdot g_{s,\text{vib},e,v}))$, with $s$ fixed, lay in the same curve, this being of slope $-1/(k_B T)$ and of $y$-intercept $\ln(x_{s}/ Q_{s,\text{el-vib}}(T))$. If the vibrational partition function $Q_{s,\text{vib},e}$ does not vary too much with the electronic level $e$ such that $Q_{s,\text{vib},e}\approx Q_{s,\text{vib}}$, one may also show from relation \eqref{eq:T_s_el_rep_2} that in thermal equilibrium the points of abscissae $\epsilon_{s,\text{el},e}$ and ordinates $\ln(x_{s,e}/g_{s,\text{el},e})$, with $s$ fixed, lay in the same curve, this being of slope $-1/(k_B T)$ and of $y$-intercept $\ln(x_{s}/Q_{s,\text{el}}(T))$.

The representative vibrational and electronic temperatures may in their turn be used to compute Boltzmann representative vibronic mole fractions $x_{s,e,v}^B:=x_{s,e,v}(T_{s,\text{vib},e})$ - given by \eqref{eq:T_s_vib_rep_2} - and Boltzmann representative electronic mole fractions $x_{s,e}^B:=x_{s,e}(T_{s,\text{vib},e},T_{s,\text{el}})$ - given by \eqref{eq:T_s_el_rep_2}.

\begin{figure}[H]
\centering
\centerline{\includegraphics[scale=1]{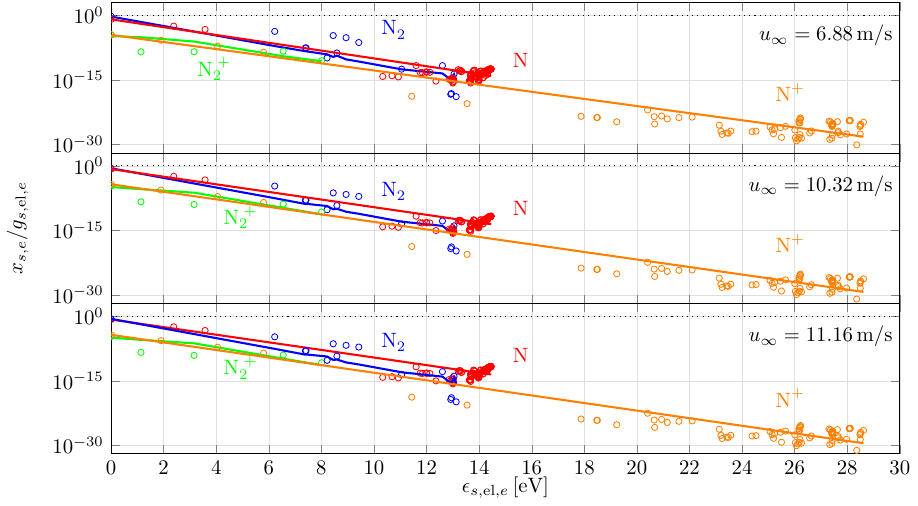}}
\vspace{-15pt}
\caption{Electronic state-specific mole fractions $x_{s,e}/g_{s,\text{el},e}$ (markers) and respective Boltzmann representatives $x_{s,e}^\text{B}/g_{s,\text{el},e}$ (lines) as functions of the electronic energies $\epsilon_{s,\text{el},e}$, obtained from the simulations of the low, medium and high speed shots, at $x=5\,\text{cm}$.}
\label{fig:Results_x_s_e__vs__T_e__far_1D}
\end{figure}

\begin{figure}[H]
\centering
\centerline{\includegraphics[scale=1]{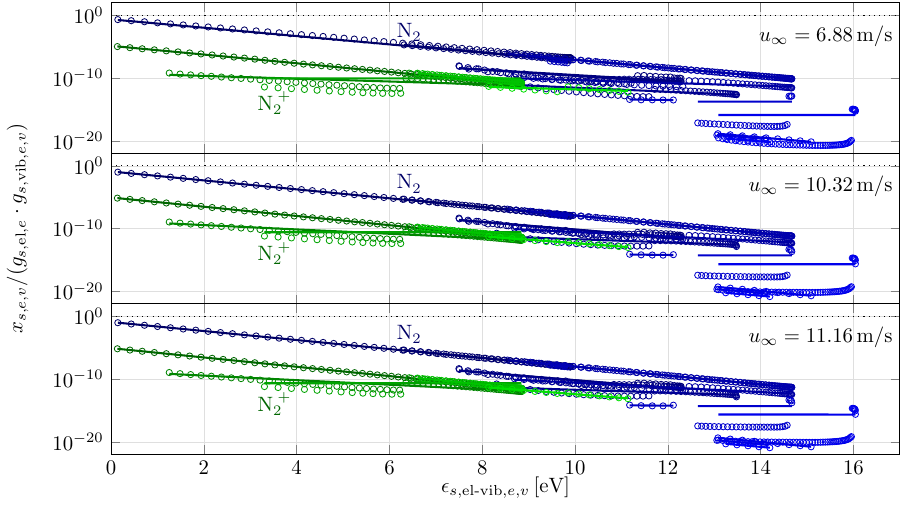}}
\vspace{-15pt}
\caption{Vibronic state-specific mole fractions $x_{s,e,v}/(g_{s,\text{el},e}\cdot g_{s,\text{vib},e,v})$ (markers) and respective Boltzmann representatives $x_{s,e,v}^\text{B}/(g_{s,\text{el},e}\cdot g_{s,\text{vib},e,v})$ (lines) as functions of the vibronic energies $\epsilon_{s,\text{el-vib},e,v}$, obtained from the simulations of the low, medium and high speed shots, at $x=5\,\text{cm}$.}
\label{fig:Results_x_s_e_v__vs__T_e_v__far_1D}
\end{figure}

Figure \ref{fig:Results_x_s_e__vs__T_e__far_1D} shows the electronic state-specific mole fractions $x_{s,e}/g_{s,\text{el},e}$ (markers) and respective Boltzmann representatives $x_{s,e}^\text{B}/g_{s,\text{el},e}$ (lines) obtained from the simulations of the low, medium and high speed shots, at $x=5\,\text{cm}$. The markers in the figure sparsely agree with the respective lines. This result shows that self-equilibrium of the electronic energy modes of the different species was not attained at $x=5\,\text{cm}$. 
The obtained representative electronic temperatures are presented by Table \ref{tab:results_T_el}. It shows that these temperatures depart from each other and from the obtained heavy particle and free electron translational ones in the order of the thousands of kelvins. In fact, the variables $T_{s,\text{el}}$ are all lower than $T_{\text{tr}_\text{h}}$, by amounts ranging from $3,359\,\text{K}$ (or $43.10\,\%$) to $1,407\,\text{K}$ (or $18.05\,\%$), $2,811\,\text{K}$ (or $39.13\,\%$) to $968\,\text{K}$ (or $13.47\,\%$), and $2740\,\text{K}$ (or $38.60\,\%$) to $932\,\text{K}$ (or $13.13\,\%$), for the cases of the low, medium and high speed shots, respectively. Therefore, one cannot state that the electronic and the translational energy modes attained an equilibrium with each other at $x=5\,\text{cm}$. 
\begin{table}[H]
\centering
\caption{Heavy particle and free electron translational temperatures $T_{\text{tr}_\text{h}}$ and $T_{\text{tr}_\text{e}}$, and representative electronic temperatures $T_{s,\text{el}}$ obtained from the simulations of the low, medium and high speed shots, at $x=5\,\text{cm}$.}
\begin{scriptsize}
\begin{tabular}{c|cccccc}
\toprule
$u_\infty[\text{m}/\text{s}]$ & $T_{\text{tr}_\text{h}}[\text{K}]$ & $T_{\text{tr}_\text{e}}[\text{K}]$ & $T_{\ch{N2},\text{el}}[\text{K}]$ & $T_{\ch{N},\text{el}}[\text{K}]$ & $T_{\ch{N2+},\text{el}}[\text{K}]$ & $T_{\ch{N+},\text{el}}[\text{K}]$\\
\midrule
$6.88$ & $7,793$ & $7,811$ & $4,434$ & $5,551$ & $6,386$ & $6,093$ \\
$10.32$ & $7,184$ & $7,187$ & $4,373$ & $5,682$ & $6,216$ & $5.762$\\
$11.16$ & $7,098$ & $7,100$ & $4,358$ & $5,699$ & $6,166$ & $5,719$\\ 
\bottomrule
\end{tabular}
\end{scriptsize}
\label{tab:results_T_el}
\end{table}
Figure \ref{fig:Results_x_s_e_v__vs__T_e_v__far_1D} shows the obtained vibronic state-specific mole fractions $x_{s,e,v}/(g_{s,\text{el},e}\cdot g_{s,\text{vib},e,v})$ and respective Boltzmann representatives $x_{s,e,v}^\text{B}/(g_{s,\text{el},e}\cdot g_{s,\text{vib},e,v})$ at $x=5\,\text{cm}$. In this figure, there are markers that agree remarkably well with the respective lines (vibrational self-equilibrium was achieved in those cases), and others (the ones associated with the seemingly horizontal lines) which do not agree at all. The deviations of the latter are due to the non-modelling of the spontaneous emission processes of the higher vibronic levels, implying these to not be depopulated through these processes in contrast to the lower ones. When fitting the respective vibronic state-specific mole fractions, a nearly horizontal line is obtained, which is associated with a very high representative vibrational temperature. This result is clearly unphysical. To avoid it in the future, one should consider a redistribution procedure on the Einstein coefficients for spontaneous emission to model the vibronic levels whose data are not available. 
\noindent
The non-absurd values of $T_{s,\text{vib},n}$ deviate from $T_{\text{tr}_\text{h}}$ by amounts from $-843\,\text{K}$ (or $-10.82\,\%$) to $3,730\,\text{K}$ (or $47.86\,\%$), $-383\,\text{K}$ (or $-5.33\,\%$) to $1,112\,\text{K}$ (or $15.48\,\%$), and $-311\,\text{K}$ (or $-4.38\,\%$) to $1,188\,\text{K}$ (or $16.74\,\%$), for the cases of the low, medium and high speed shots, respectively. These values are too large for one to assume that equilibrium between the vibrational and translational energy modes was attained for all the considered electronic levels. It is important to note that markers associated with the same species do not lay on the same line, showing that the respective electronic energy mode is not at self-equilibirum (as also evidenced by the fact that the markers in Figure \ref{fig:Results_x_s_e__vs__T_e__far_1D} do not coincide with their Boltzmann representatives). 

\begin{figure}[H]
\centering
\centerline{\includegraphics[scale=1]{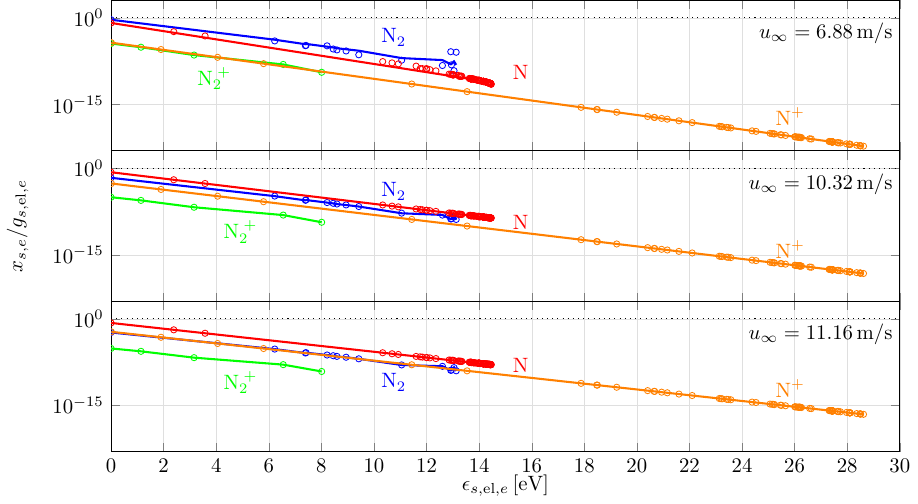}}
\vspace{-15pt}
\caption{Electronic state-specific mole fractions $x_{s,e}/g_{s,\text{el},e}$ (markers) and respective Boltzmann representatives $x_{s,e}^\text{B}/g_{s,\text{el},e}$ (lines) as functions of the electronic energies $\epsilon_{s,\text{el},e}$, obtained from the simulations of the low, medium and high speed shots, at $x=5\,\text{cm}$, disregarding spontaneous emission processes.}
\label{fig:Results_x_s_e__vs__T_e__far__1D_no_se}
\end{figure}

Since absorption and induced emission processes were not considered in this work, spontaneous emission occurred without any counterbalance. This may have led to the observed self-non-equilibrium of the electronic energy modes. Figure \ref{fig:Results_x_s_e__vs__T_e__far__1D_no_se} shows the electronic state-specific mole fractions $x_{s,e}/g_{s,\text{el},e}$ and respective Boltzmann representatives $x_{s,e}^\text{B}/g_{s,\text{el},e}$ obtained from the simulations of the low, medium and high speed shots, at $x=5\,\text{cm}$, disregarding spontaneous emission. The markers in this figure agree much better with the respective lines than in Figure \ref{fig:Results_x_s_e__vs__T_e__far_1D}. Self-equilibrium of the electronic energy modes seems to be attained by all species with the exception of \ch{N2} and \ch{N} in the case of the low speed shot, as the respective markers slightly depart from the lines. Table \ref{tab:results_T_el_no_se} presents the obtained representative electronic temperatures. The variables $T_{s,\text{el}}$ depart from $T_{\text{tr}_\text{h}}$, by amounts ranging from $-948\,\text{K}$ (or $-11,76\,\%$) to $432\,\text{K}$ (or $5.36\,\%$), $-83\,\text{K}$ (or $-0.90\,\%$) to $39\,\text{K}$ (or $0.42\,\%$), and $-81\,\text{K}$ (or $-0.80\,\%$) to $11\,\text{K}$ (or $0.11\,\%$), for the cases of the low, medium and high speed shots, respectively. All these values are negligibly small with the exception of the ones from the former case, the lower limit being due to \ch{N} and the higher limit due to \ch{N2}, which, as mentioned above, were also found to not be at electronic self-equilibrium. One may state that the electronic and the translational energy modes of the particles attained an equilibrium between themselves at $x=5\,\text{cm}$, with the exception of \ch{N} and \ch{N2} in the case of the low speed shot. 
\begin{table}[H]
\centering
\caption{Heavy particle and free electron translational temperatures $T_{\text{tr}_\text{h}}$ and $T_{\text{tr}_\text{e}}$, and representative electronic temperatures $T_{s,\text{el}}$ obtained from the simulations of the low, medium and high speed shots, at $x=5\,\text{cm}$, disregarding spontaneous emission.}
\begin{scriptsize}
\begin{tabular}{c|cccccc}
\toprule
$u_\infty[\text{m}/\text{s}]$ & $T_{\text{tr}_\text{h}}[\text{K}]$ & $T_{\text{tr}_\text{e}}[\text{K}]$ & $T_{\ch{N2},\text{el}}[\text{K}]$ & $T_{\ch{N},\text{el}}[\text{K}]$ & $T_{\ch{N2+},\text{el}}[\text{K}]$ & $T_{\ch{N+},\text{el}}[\text{K}]$\\
\midrule
$6.88$ & $8,059$ & $8,078$ & $8,491$ & $7,111$ & $8,100$ & $8.063$\\
$10.32$ & $9,264$ & $9,271$ & $9,181$ & $9.216$ & $9.303$ & $9.271$\\
$11.16$ & $10,114$ & $10,116$ & $10,033$ & $10,113$ & $10,125$ & $10,116$\\ 
\bottomrule
\end{tabular}
\end{scriptsize}
\label{tab:results_T_el_no_se}
\end{table} 

\begin{figure}[H]
\centering
\centerline{\includegraphics[scale=1]{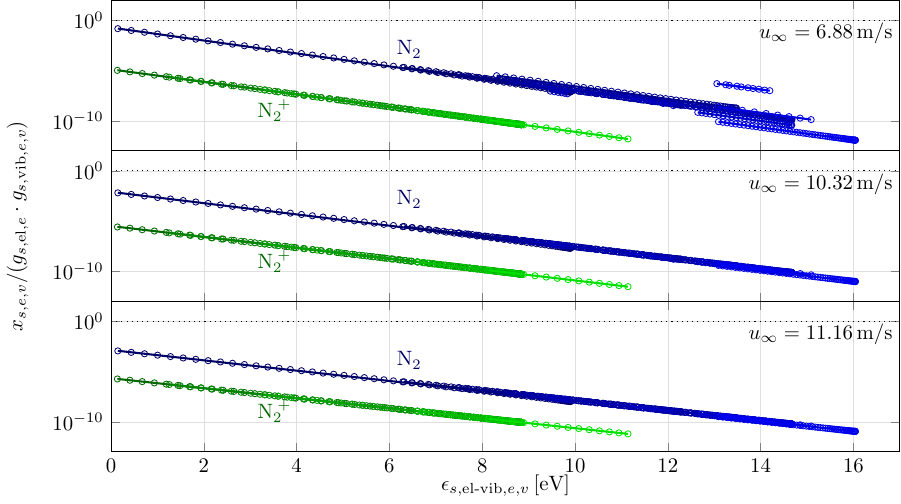}}
\vspace{-15pt}
\caption{Vibronic state-specific mole fractions $x_{s,e,v}/(g_{s,\text{el},e}\cdot g_{s,\text{vib},e,v})$ (markers) and respective Boltzmann representatives $x_{s,e,v}^\text{B}/(g_{s,\text{el},e}\cdot g_{s,\text{vib},e,v})$ (lines) as functions of the vibronic energies $\epsilon_{s,\text{el-vib},e,v}$, obtained from the simulations of the low, medium and high speed shots, at $x=5\,\text{cm}$, disregarding spontaneous emission processes.}
\label{fig:Results_x_s_e_v__vs__T_e_v__far__1D_no_se}
\end{figure}

Figure \ref{fig:Results_x_s_e_v__vs__T_e_v__far__1D_no_se} shows the vibronic state-specific mole fractions $x_{s,e,v}/(g_{s,\text{el},e}\cdot g_{s,\text{vib},e,v})$ and respective Boltzmann representatives $x_{s,e,v}^\text{B}/(g_{s,\text{el},e}\cdot g_{s,\text{vib},e,v})$, obtained from the simulations of the low, medium and high speed shots, at $x=5\,\text{cm}$, disregarding spontaneous emission. All markers in \Cref{fig:Results_x_s_e_v__vs__T_e_v__far__1D_no_se} lay in their respective lines, evidencing that vibrational self-equilibrium occurs. However, the lines associated with \ch{N2} for the case of the low speed shot depart slightly from each other. This is a result of an electronic self-non-equilibrium of \ch{N2}, which was already mentioned. The values of $T_{s,\text{vib},n}$ deviate from $T_{\text{tr}_\text{h}}$ by amounts from $-1,557\,\text{K}$ (or $-19.32\,\%$) to $46\,\text{K}$ (or $0.57\,\%$), $-1,052\,\text{K}$ (or $-11.36\,\%$) to $42\,\text{K}$ (or $0.45\,\%$), and $-729\,\text{K}$ (or $-7.21\,\%$) to $13\,\text{K}$ (or $0.13\,\%$), for the cases of the low, medium and high speed shots, respectively. The lower limits of the deviations are significant and in the same order of magnitude as the ones obtained when considering spontaneous emission. The upper limits are, however, much smaller. The vibrational and translational energy modes should therefore be closer to an equilibrium between each other when disregarding spontaneous emission than when regarding.

All of the above-mentioned results allow the authors to conclude that the spontaneous emission processes strongly contribute to the observed thermal non-equilibrium conditions of the systems at $x=5\,\text{cm}$. It should be ascertained in the future if the inclusion of absorption and induced emission processes may reduce this thermal non-equilibrium. The results also showed that collisional processes, alone, do not ensure equilibrium between the electronic and translational energy modes of \ch{N2} and \ch{N} in the case of the low speed shot.

\subsection{Dependence on the escape factor}
\label{subsection:Escape_factor_VUV}

It is well known that the emitted  VUV radiation is strongly absorbed by the particles in the medium \cite{cruden2019} and, therefore, the assumption of it being optically thin may not be valid. By considering an escape factor $\Lambda^{\text{VUV}}<1$ in the simulations, it is possible to take into account, in a crude way, the effect of auto-absorption (i.e. absorption by the same emitting source) of the VUV radiation on the resultant radiation variables. Part of the energy which was once lost due to emitted VUV radiation escaping from the system may now be used in other processes. More energy becomes available for excitation of the particles, and as a consequence, more energy is radiated in the other wavelength intervals. There is even the possibility of more radiative energy in the VUV wavelength interval escaping from the system, since though a significant fraction of the photons do not escape, the number of emitted photons becomes as high as the number excited particles. Values of $\Lambda^{\text{VUV}}=0.1$ and $0.01$ were tried, and the respective overall radiation variables are shown in \Cref{fig:Results_I_inst_VUV_1D_LL,fig:Results_I_ll_ne_inst_VUV_1D_LL,fig:Results_I_inst_Blue_1D_LL,fig:Results_I_ll_ne_inst_Blue_1D_LL,fig:Results_I_inst_Red_1D_LL,fig:Results_I_ll_ne_inst_Red_1D_LL,fig:Results_I_inst_IR_1D_LL,fig:Results_I_ll_ne_inst_IR_1D_LL} in \ref{sec:appendixA}. 
Figure \ref{fig:Results_I_inst_VUV_1D_LL} shows that by decreasing the escape factor, the peak values of the instrumentally resolved radiative intensities in the VUV wavelength region decreased as much as a half in the three shots. On the other hand, the contributions of \ch{N} increased significantly becoming now much more dominant than the others (i.e. the ones of $\ch{N2+}(\text{C}-\text{X})$, $\ch{N2}(\text{a}-\text{X})$ and $\ch{N2}(\text{A}-\text{X})$), as shown in Figure \ref{fig:Results_I_ll_ne_inst_VUV_1D_LL}). The obtained spectra show better agreement with the experimental results. The radiative intensity profiles show steeper rises and falls, and, in the case of the low speed shot, get closer to the experimental ones. Still, no plateaus were predicted in the cases of the medium and high speed shots. For the case of the ``Blue'' wavelength region, the decrease in the escape factor made the peak values of the instrumentally resolved radiative intensities to increase with some significance: as much as 10\% in the low speed shot, and 40\% in the medium and high speed shots. The change on the peaks shapes is not noticeable (see Figure \ref{fig:Results_I_inst_Blue_1D_LL}). Regarding the shape of the instrumentally resolved non-equilibrium metric, Figure \ref{fig:Results_I_ll_ne_inst_Blue_1D_LL} shows that only the peak associated with $\ch{N2}(\text{C}-\text{B})$ at $\lambda=337\,\text{nm}$ experienced a meaningful change, increasing its height with the decrease in the escape factor value. For the case of the ``Red'' and IR wavelength regions, the  peak values of the instrumentally resolved radiative intensities also increased: as much as 10\% in the low speed shot, and 60\% (for the case of the former) and 80\% (for the case of the latter) in the medium and high speed shots. The shapes of the peaks did not change significantly. The contributions from \ch{N} to the non-equilibrium metrics were preferentially augmented (in particular in the medium and high speed shots), agreeing better with the experimental results.

These results show strong evidence for the medium being optically thick in the VUV wavelength region.

\subsection{Dependence on the dissociation rates of \ch{N2}}
\label{subsection:VD_dependence}

The evolution of the post-shock conditions is highly dependent on the dissociation rates of \ch{N2}. Greater dissociation rates would imply a production of a greater amount of \ch{N} atoms. And, due to dissociation being an endothermic process, lesser energy would be available for excitation. However, it is hard to say how an increase of the dissociation rates could affect the excitation of the particles to a given energy level since a lesser population of \ch{N2} would mean a lower number of \ch{N2} molecules to be excited and the converse for the case of \ch{N}. As such, one may not say, in a straightforward manner if the number of particles in some excited energy level should be lower or higher. 
To test the dependence of the numerical results on the dissociation rates of \ch{N2}, we decided to consider dissociation rates of $\ch{N2}(\text{X}{}^1\Sigma_\text{g}^+)$ (the ground electronic level of molecular nitrogen\footnote{The ground electronic level of \ch{N2} is usually more populated than the other ones, being, therefore, more important in what concerns dissociation.}) scaled by $0.1$ and $10$, keeping $\Lambda^{\text{VUV}}=0.01$. The obtained results are shown in \Cref{fig:Results_I_inst_VUV_1D_VD,fig:Results_I_ll_ne_inst_VUV_1D_VD,fig:Results_I_inst_Blue_1D_VD,fig:Results_I_ll_ne_inst_Blue_1D_VD,fig:Results_I_inst_Red_1D_VD,fig:Results_I_ll_ne_inst_Red_1D_VD,fig:Results_I_inst_IR_1D_VD,fig:Results_I_ll_ne_inst_IR_1D_VD} in \ref{sec:appendixA}. These show that the peak values of the instrumentally resolved radiative intensities increased with a decrease of the rates of dissociation with this being more relevant for the case of all wavelength intervals for the low speed shot and the wavelength intervals ``Blue'', ``Red'' and IR for the medium and high speed shots. The increase went from insignificant to as high as $100\,\%$. For the case of an increase of the dissociation rates, the decrease of the peak values went from as low as $20\,\%$ to as high as $70\,\%$. The figures for the non-equilibrium metrics show that both contributions of \ch{N} and \ch{N2} to the radiation variables increased with the decrease of the rates of dissociation, with the effect on the latter contribution being much more significant. The energy that was once spent in dissociation was instead spent on the excitation of \ch{N} and \ch{N2}, and since more \ch{N2} and less \ch{N} particles were obtained, the effect on \ch{N2} had a greater importance. Note, however, that to get a better agreement with the experimental spectra, the contributions of \ch{N} should be the ones to get enhanced and not the contributions of \ch{N2}. Also, one should point out that in the case of the low speed shot, the radiative intensity peaks widened with the decrease of the rates of dissociation, agreeing better with the experimental profiles for all wavelength intervals except for VUV. For the other shots, the change on the peaks shape was not significant. The increase of the dissociation rates made the peaks to narrow in the case of the low speed shot, agreeing worse with experimental counterparts. Conversely, the increase widened the peaks in the case of the ``Blue'' wavelength interval of the medium and high speed shots, getting closer to the experimental results.

In short, solely decreasing or increasing the dissociation rates of $\ch{N2}(\text{X})$ do not unequivocally lead to better or worse results.

\subsection{Dependence on the excitation rates of \ch{N}}
\label{subsection:Excitation_of_N}

It was mentioned in the section that precedes the above one that by considering an escape factor $\Lambda^{\text{VUV}}<1$ the contributions of \ch{N} to the radiation variables were still underestimated when compared with the molecular contributions in the ``Red'' and IR wavelength intervals. It was then decided to try scaling the excitation rate coefficients of \ch{N}, keeping $\Lambda^{\text{VUV}}=0.01$, to enhance them. The crude formulae of Annaloro \textit{et al.} \cite{annaloro2014a} and Panesi \textit{et al.} \cite{panesi2009} regarded in the database\footnote{ See companion paper CITE COMPANION PAPER.} as models for excitation of \ch{N} by heavy particle and electron impact, respectively, may indeed accommodate some degree of uncertainty. Their multiplication by $10$ and $100$ was tried, and numerical simulations solely converged for the case of low speed shot. This is an indication of the rate coefficients being actually high enough in the conditions of the medium and high speed shots, wherein increasing their values may lead to a physical incoherence. The results are shown in \Cref{fig:Results_I_inst_1D_LL_E_N,fig:Results_I_ll_ne_inst_1D_LL_E_N} in \ref{sec:appendixA}. 
With the increase in the rate coefficients for excitation of \ch{N}, the peaks values of the  instrumentally resolved radiative intensities rose as much as to the quintuple in the case of the VUV radiation, the double in the case of the ``Blue'' and ``Red'' radiation, and the quadruple in the case of the IR radiation. However, the values are still one to two orders of magnitude lower than the experimental ones. Regarding the shape of the profiles, it was found that the rising and falling parts got steeper, deviating significantly from the experimental profiles. This result may be justified by a greater rate excitation rate of \ch{N} producing a greater rate of radiative emission, and, therefore, the radiative intensity rises faster and higher. As a higher excitation requires more energy, which is then lost in the form of radiation, the system suddenly gets incapable of continually exciting the particles, and the radiative intensity falls faster. \Cref{fig:Results_I_ll_ne_inst_1D_LL_E_N} shows that the contribution of \ch{N} to the instrumentally resolved non-equilibrium metrics increased in all wavelength regions, agreeing better with the experimental spectra except in the ``Blue'' wavelength region, for which the contributions of the molecular particles should prevail over the ones of the atomic particles. 

In short, the increase of the rate coefficients for excitation of \ch{N} cannot be said to unequivocally lead to better results.

\subsection{A synopsis about the dependence of the results on the different parameters}
\label{subsection:Dependence_synopsis}

\begin{figure}[H]
\centering
\centerline{\includegraphics[scale=1]{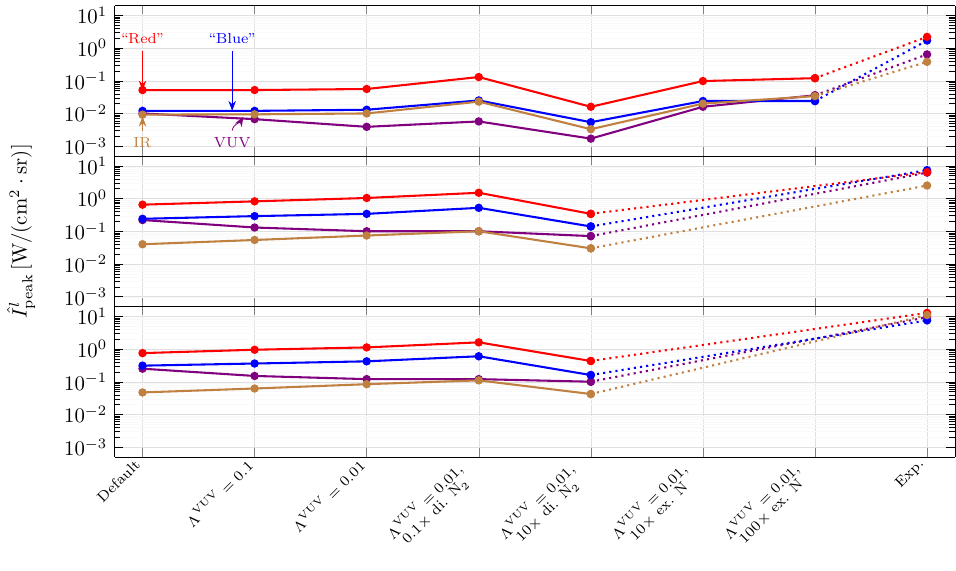}}
\vspace{-15pt}
\caption{Peak values of the instrumentally resolved radiative intensities $\hat{I}^{l}_{\text{peak}}$ obtained with the different models and in the experiment for the low, medium and high speed shots.}
\label{fig:Results_I_inst_peak_synopsis}
\end{figure}

\begin{figure}[H]
\centering
\centerline{\includegraphics[scale=1]{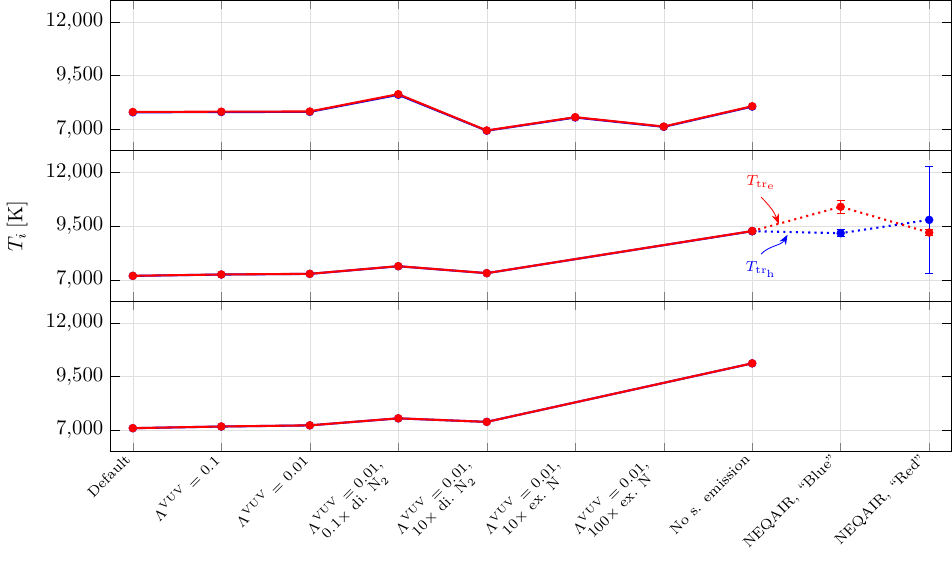}}
\vspace{-15pt}
\caption{Temperatures $T_{\text{tr}_\text{h}}$ and $T_{\text{tr}_\text{e}}$ at $x=5\,\text{cm}$ obtained with the different models for the low, medium and high speed shots.}
\label{fig:Results_T_synopsis}
\end{figure}

\begin{figure}[H]
\centering
\centerline{\includegraphics[scale=1]{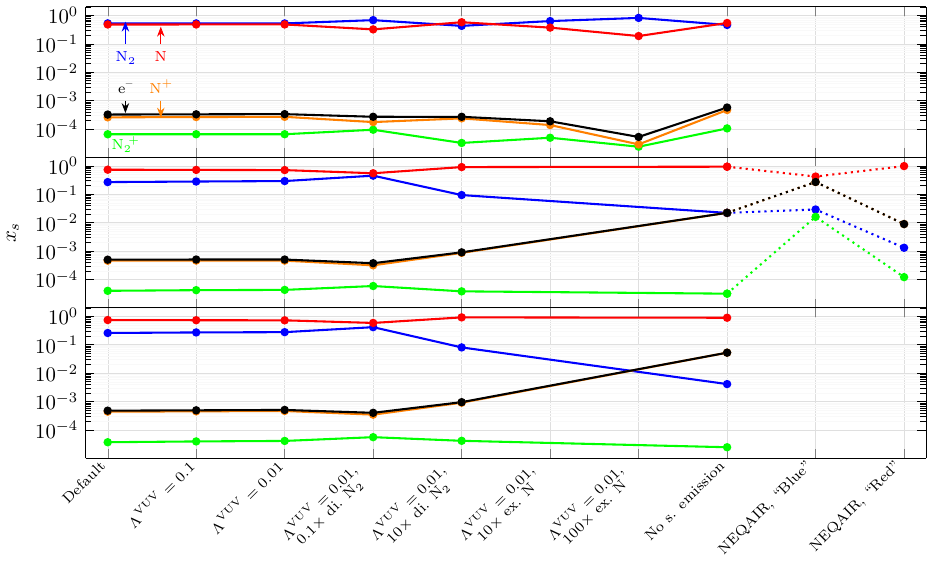}}
\vspace{-15pt}
\caption{Mole fractions $x_s$ at $x=5\,\text{cm}$ obtained with the different models for the low, medium and high speed shots.}
\label{fig:Results_x_s_synopsis}
\end{figure}

\Cref{fig:Results_I_inst_peak_synopsis} shows the peak values of the instrumentally resolved radiative intensities $\hat{I}^{l}_{\text{peak}}$ (with $l\in\{\text{VUV}$,$ \text{``Blue''}$,$ \text{``Red''}$,$\text{IR}\}$) obtained with the different models (i.e. the default model and its transformations) and in the experiment for the low, medium and high speed shots. And \Cref{fig:Results_T_synopsis,fig:Results_x_s_synopsis} show the temperatures $T_{\text{tr}_\text{h}}$ and $T_{\text{tr}_\text{e}}$ and mole fractions $x_s$ attained at $x=5\,\text{cm}$, respectively, which were obtained with the different models of this work and of the work of Cruden and Brandis \cite{cruden2019}.

By considering escape factors $\Lambda^{\text{VUV}}=0.1$ and $\Lambda^{\text{VUV}}=0.01$ for the VUV wavelength interval, the peak values of the instrumentally resolved radiative intensities increase for all wavelength intervals except VUV, for which it decreases. The attained temperatures rise just slightly, but the mole fractions of the species do not change appreciably.

Decreasing the rates of dissociation of $\ch{N2}(\text{X}{}^1\Sigma_\text{g}^+)$, keeping $\Lambda^{\text{VUV}}=0.01$, makes the peak values of the instrumentally resolved radiative intensities to increase, and increasing the rates of dissociations makes them to decrease. The impact is, however, not so relevant for the case of the VUV wavelength interval of the medium and high speed shots. \Cref{fig:Results_T_synopsis} shows that by decreasing the rates of dissociation, less energy is spent on endothermic processes (since higher translational temperatures are attained). Curiously, increasing the rates of dissociation does not decrease the attained  translational temperatures in the cases of the medium and high speed shots (the translational energy is then redistributed in a different way). It is important to mention here that the values of all of the referred temperatures are still much lower (by several thousands of kelvins) than the ones inferred by Cruden and Brandis. \Cref{fig:Results_x_s_synopsis} shows a tendency for an increase of the attained mole fractions of \ch{N} and \ch{N+} and a decrease for the ones of \ch{N2} and \ch{N2+} with an increase of the rates of dissociation. The increase on the mole fraction of \ch{N+} and the decrease on the one of \ch{N2+} may be justified by the fact that a stronger dissociation implies that, in contrast to \ch{N2}, more \ch{N} particles become available to be ionised. The values for the mole fractions obtained by increasing the rate coefficients are actually the ones of all the tried models (regarding spontaneous emission) which better agree with the ones of Cruden and Brandis. Still, their results indicate that the system should have endured stronger dissociation and ionisation.

By keeping $\Lambda^{\text{VUV}}=0.01$, and increasing the excitation rates of \ch{N} (which was only successful for the case of the low speed shot), a meaningful rise of the peak radiation values occurs, in particular the ones for the IR and VUV wavelength intervals. Still, these are one to two orders of magnitude lower than the ones obtained in the experiment (labelled by ``Exp.''). The increase in the radiation variables comes at a cost of the translational temperatures ($T_{\text{tr}_\text{h}}$ decreases as much as $698\,\text{K}$) and a weaker dissociation and ionisation of \ch{N2}. The mole fraction of \ch{N+} at $x=5\,\text{cm}$ gets lower due to a faster recombination. The peak value of the mole fraction of \ch{N+} actually increases as the highly excited \ch{N} particles are more easily ionised. It can be concluded that more energy is used in endothermic processes, with excitation of \ch{N} being the mainly one.

As shown by \Cref{fig:Results_T_synopsis,fig:Results_x_s_synopsis}, the translational temperatures and the mole fractions of all species except \ch{N2+} attained in the numerical simulations disregarding spontaneous emission (labelled by ``No s. emission'') agree well with the ones derived by Cruden and Brandis. This endorses the hypothesis of the simulated elements of fluid not receiving the amount of energy that they should, as only by disregarding spontaneous emission (therefore, retaining a lot more energy) the thermodynamic conditions get reasonably closer to the inferred ones.

\subsection{Possible causes of the significant underestimation of the experimental results by the numerical ones}
\label{subsection:causes_understimation}

Several studies in the literature reported the implementation of analytical and numerical models which significantly underestimated the radiation variables obtained trough shock tube experiments. In fact, these discrepancies had been so recurrently observed that Cruden \textit{et al.} \cite{cruden2015} even used the term ``long-standing'' to describe the problem regarding results obtained from the Japan Aerospace Exploration Agency's High-Enthalpy Shock Tunnel (JAXA-HIEST). Cruden \textit{et al.} \cite{cruden2015} studied the impact that contamination species have on the numerical results, fitting their concentrations to match some particular JAXA-HIEST experimental results. They concluded that it may well be possible for contaminants such as \ch{Fe} (atomic iron) and \ch{CN} (cyanogen radical) to cause the observed discrepancies. In the case of the experiments that produced the benchmark data analysed in the present work - the EAST test 62 - Cruden and Brandis \cite{cruden2019} found evidence for the presence of the contaminants \ch{C} (atomic carbon), \ch{H} (atomic hydrogen) and \ch{CN}(cyanogen radical). However, it seems improbable that this could explain the deviations of several orders of magnitude for the obtained radiation variables, since the observed spectrum is almost completely dominated by non-contaminant species (namely \ch{N}, \ch{N2} and \ch{N2+}). Furthermore, Brandis \textit{et al.} \cite{brandis2010} refer that in the latest EAST campaigns many upgrades to the system have been made to reduce the level of contamination.

A more feasible contribution to the discrepancies between the numerical and experimental results is the non-modelling of the so-called precursor phenomena: some of the VUV radiation emitted by highly excited particles in the shock layer is absorbed by others upstream of the shock wave, inducing their excitation, photoionisation and photodissociation. This changes the conditions upstream of the shock wave and, inevitably, the conditions downstream of it: Nomura \textit{et al.} \cite{nomura2020} refer that the shock layer thickness and the non-equilibrium temperatures are increased, yielding an excess of radiation. Yamada \textit{et al.} \cite{yamada2019} compared experimental results with numerical ones obtained through a model that does not account for precursor phenomena, and found that the measured radiation intensities for \ch{N2}, \ch{N2+} and \ch{N} started to increase upstream of the shock wave. The radiation profiles in the shock layer differed significantly, showing that the precursor phenomena had a great influence on the thermochemical processes that occurred downstream of the shock wave.

Another feasible contribution to the discrepancies corresponds to heating of the driven gas due to downstream plasma subjected to a stronger shock wave, and radiative energy transfer from the driver gas and the EAST electric arc. Bogdanoff and Park \cite{bogdanoff2002} performed several shock tube experiments, finding the temperature downstream of the shock wave in the observation point to be three to four times the one obtained through the Rankine-Hugoniot relations. The electric arc of the shock tube increases the temperature of the driver gas to very high values (several tens of thousands of kelvins) which causes it to radiate a lot of energy, some of it to the driven gas upstream and downstream of the shock wave. Also, it is known that the shock wave decelerates through the shock tube, heating more the part of the driven gas near the diaphragm than the part of the driven gas near the observation point\footnote{In a correspondence, Dr. Brett Cruden stated that this contribution was suspected to be the most important one.}. A transfer of energy from the former to the later may then also occur.

\section{Conclusions}
\label{sec:conclusions}

The database of kinetic processes described in {\color{red} CITE COMPANION PAPER} was employed in Euler one-dimensional simulations of the shots $19$, $20$ and $40$ of the EAST's $62^{\text{nd}}$ campaign. It was found that the model underpredicted the experimental radiation variables by one to two orders of magnitude. Sensitivity tests performed on different parameters of the simulations were unsuccessful in getting a reasonable agreement. The shape of the radiative intensities profiles of the low speed shot was correctly predicted, but not the ones of the medium and high speed shots which revealed non-null plateaus proceeding or coalescing with peaks. These plateaus were not predicted at all. The analysis of Cruden and Brandis on the spectra obtained in the shot $19$  of the EAST's $62^{\text{nd}}$ campaign showed that the high experimental values of the radiation variables were attained with a lower cost of translation temperature. There is a strong evidence for the underestimation of the radiation variables observed in this work to result from several contributions already pointed out by other researchers in the literature: the non-modelling of the precursor phenomena and/or the non-modelling of shock tube-related phenomena such as heat transfer by radiation between the driver gas (as well as the driver arc) and the test gas, and by conduction due to downstream plasma subjected to a stronger shock wave.

One of the next steps that should be taken in the future is to test other kinetic databases reported in the literature and to compare their results with the ones of this work, such that particular qualities which were  actually improved or, on the contrary, worsened may be identified. To quantify the possible divergences between the herein implemented vibronic-specific state-to-state model and the simpler models such as the multi-temperature ones, these latter should also be tried.

To ascertain the effect of heat transfer within the test gas subjected to strong shock waves generated by shock tubes, higher fidelity one-dimensional simulations should be performed. The crude concept of an escape factor should be disregarded, and an equation of radiative transfer should be solved instead. Furthermore, the transport phenomena should be introduced in the balance equations. 

The database of radiative processes should be extended, accommodating absorption, induced emission, photodissociation, photoassociation, photoionisation, photorecombination and bremsstrahlung, beyond spontaneous emission processes. All these processes should be treated as rovibronic instead of simply vibronic since the numerical spectra associated with the molecular contributions obtained in this work showed pointier profiles when compared to the ones obtained in the experiments.

An extrapolation of the Einstein coefficients for spontaneous emission to the energy levels whose data are not available should be performed to avoid getting unphysical levels' population distributions.

Finally, as more ambitious goals, the impact on the radiation variables of the precursor phenomena, the absorption of radiation emitted by the driver gas and the EAST electric arc, and the conduction of heat due to downstream plasma being subjected to a stronger shock wave, should be studied.

\begin{acknowledgement}

%

The authors would like to acknowledge Bruno Lopez from the University of Illinois Urbana--Champaign who provided technical support for the SPARK code, Daniel Potter from CSIRO in Australia for insights on line-broadening mechanisms, and Brett A. Cruden and Aaron Brandis from the NASA Ames Research Center, who shared their obtained data regarding the EAST's $62^{\text{nd}}$ campaign.

This work has been partially supported by the Portuguese Science Foundation FCT, under Projects UIDB/50010/2020 and UIDP/50010/2020.
 
\end{acknowledgement}


\bibliography{bibliography}

\newpage

\appendix

\section{Appendix A: Sensitivity tests on different parameters of the simulations}
\label{chapter:Sensibility_tests}
\setlabel{Appendix A}{sec:appendixA}

This appendix presents the numerical instrumentally resolved radiative intensities $\hat{I}^{l}(x)$ and non-equilibrium metrics $\hat{I}_{\lambda}^{\,\text{ne},l}(\lambda)$, with $l\in\{\text{VUV},\text{``Blue''},\text{``Red''},\text{IR}\}$, obtained from the numerical simulations of the shots 40, 19, and 20 of the $62^{\text{nd}}$ EAST campaign \cite{brandis2018shock}, considering changes of scale of the VUV escape factor $\Lambda^{\text{VUV}}$, the dissociation rates of $\ch{N2}(\text{X})$, and the excitation rates of \ch{N}. 

\subsection{Dependence on the escape factor}
\label{section:appendix_Escape_factor_VUV}

\begin{figure}[H]
\centering
\centerline{\includegraphics[scale=1]{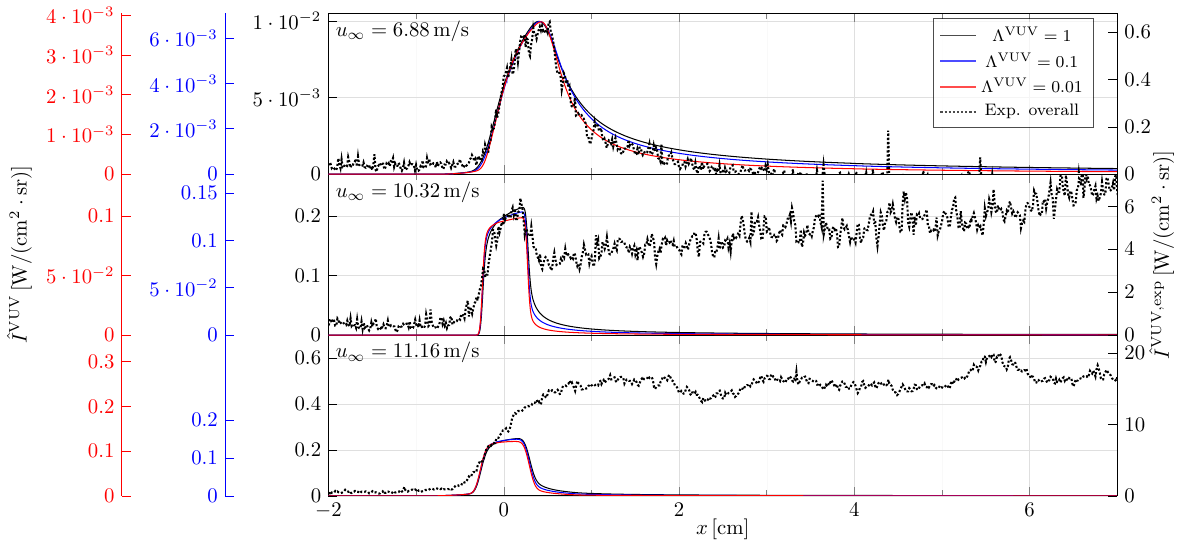}}
\vspace{-15pt}
\caption{Numerical instrumentally resolved radiative intensities $\hat{I}^{\text{VUV}}(x)$, obtained with $\Lambda^{\text{VUV}}=1$ (solid black lines), $\Lambda^{\text{VUV}}=0.1$ (solid blue lines), and $\Lambda^{\text{VUV}}=0.01$ (solid red lines), as well as the respective experimental instrumentally resolved radiative intensities $\hat{I}^{\text{VUV},\text{exp}}(x)$ (dotted black lines).}
\label{fig:Results_I_inst_VUV_1D_LL}
\end{figure}
\begin{figure}[H]
\centering
\centerline{\includegraphics[scale=1]{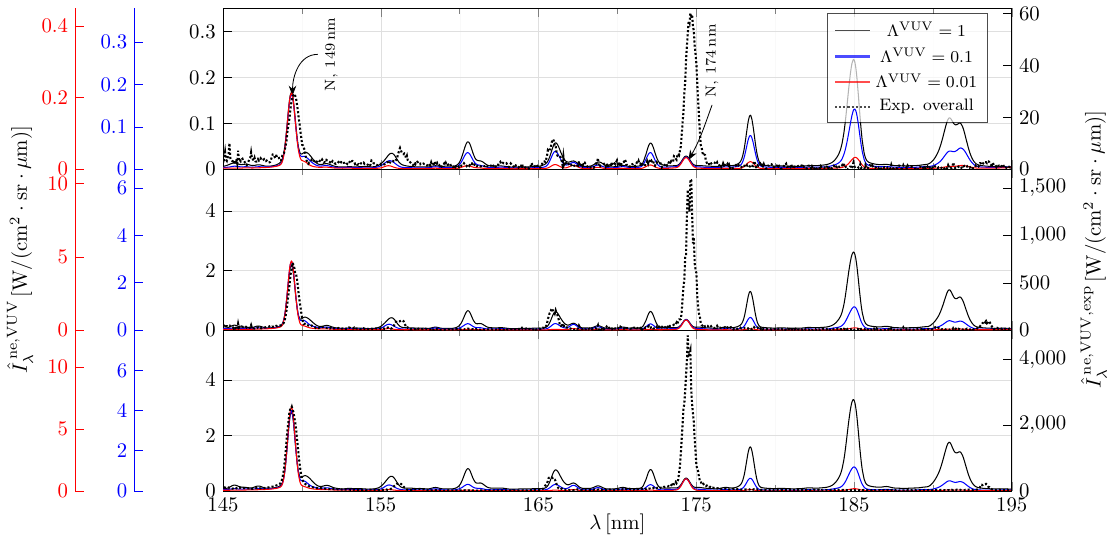}}
\vspace{-15pt}
\caption{Numerical instrumentally resolved non-equilibrium metrics $\hat{I}_{\lambda}^{\,\text{ne},\text{VUV}}(\lambda)$, obtained with $\Lambda^{\text{VUV}}=1$ (solid black lines), $\Lambda^{\text{VUV}}=0.1$ (solid blue lines), and $\Lambda^{\text{VUV}}=0.01$ (solid red lines), as well as the respective experimental instrumentally resolved non-equilibrium metrics $\hat{I}_{\lambda}^{\,\text{ne},\text{VUV},\text{exp}}(\lambda)$ (dotted black lines).}
\label{fig:Results_I_ll_ne_inst_VUV_1D_LL}
\end{figure}

\begin{figure}[H]
\centering
\centerline{\includegraphics[scale=1]{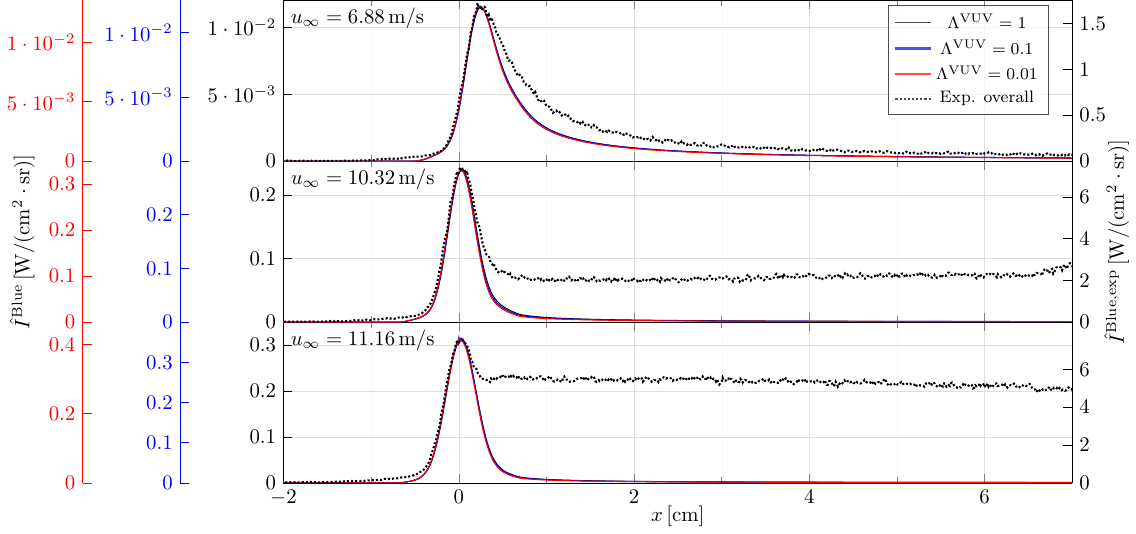}}
\vspace{-15pt}
\caption{Numerical instrumentally resolved radiative intensities $\hat{I}^{\text{Blue}}(x)$, obtained with $\Lambda^{\text{VUV}}=1$ (solid black lines), $\Lambda^{\text{VUV}}=0.1$ (solid blue lines), and $\Lambda^{\text{VUV}}=0.01$ (solid red lines), as well as the respective experimental instrumentally resolved radiative intensities $\hat{I}^{\text{Blue},\text{exp}}(x)$ (dotted black lines).}
\label{fig:Results_I_inst_Blue_1D_LL}
\end{figure}
\begin{figure}[H]
\centering
\centerline{\includegraphics[scale=1]{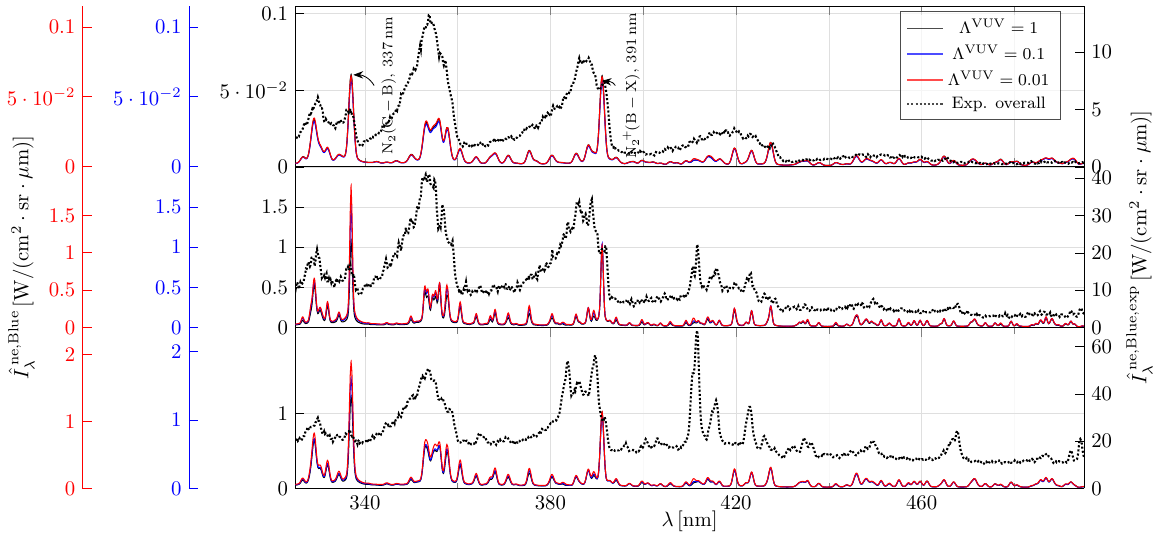}}
\vspace{-15pt}
\caption{Numerical instrumentally resolved non-equilibrium metrics $\hat{I}_{\lambda}^{\,\text{ne},\text{Blue}}(\lambda)$, obtained with $\Lambda^{\text{VUV}}=1$ (solid black lines), $\Lambda^{\text{VUV}}=0.1$ (solid blue lines), and $\Lambda^{\text{VUV}}=0.01$ (solid red lines), as well as the respective experimental instrumentally resolved non-equilibrium metrics $\hat{I}_{\lambda}^{\,\text{ne},\text{Blue},\text{exp}}(\lambda)$ (dotted black lines).}
\label{fig:Results_I_ll_ne_inst_Blue_1D_LL}
\end{figure}

\begin{figure}[H]
\centering
\centerline{\includegraphics[scale=1]{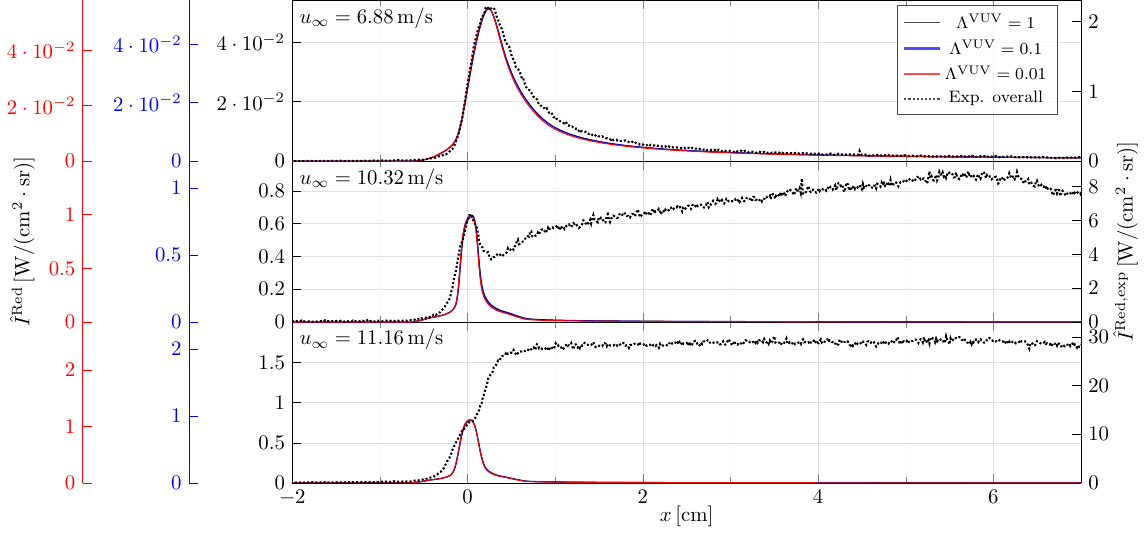}}
\vspace{-15pt}
\caption{Numerical instrumentally resolved radiative intensities $\hat{I}^{\text{Red}}(x)$, obtained with $\Lambda^{\text{VUV}}=1$ (solid black lines), $\Lambda^{\text{VUV}}=0.1$ (solid blue lines), and $\Lambda^{\text{VUV}}=0.01$ (solid red lines), as well as the respective experimental instrumentally resolved radiative intensities $\hat{I}^{\text{Red},\text{exp}}(x)$ (dotted black lines).}
\label{fig:Results_I_inst_Red_1D_LL}
\end{figure}
\begin{figure}[H]
\centering
\centerline{\includegraphics[scale=1]{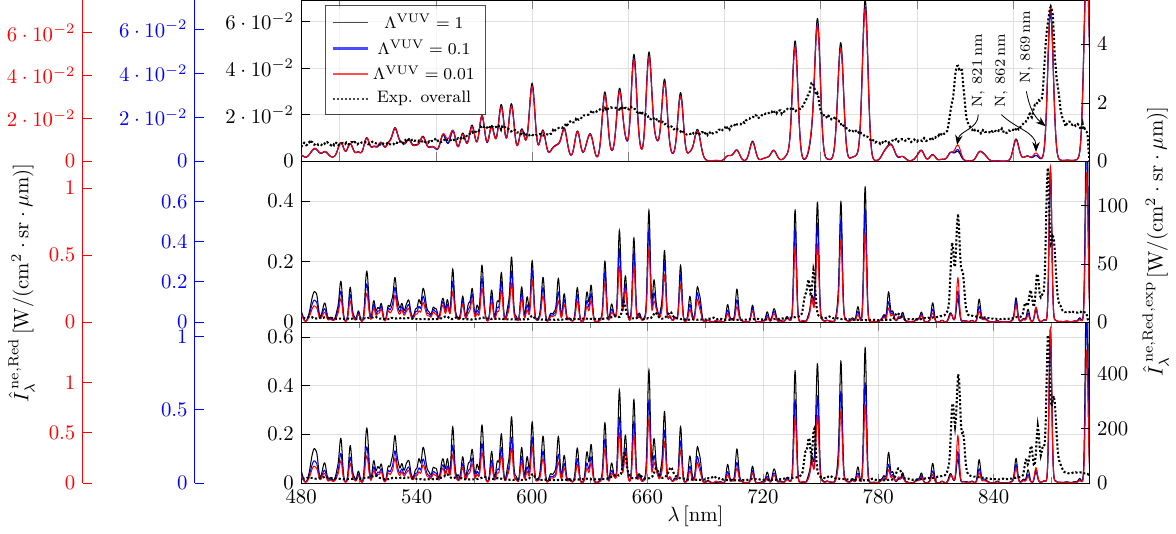}}
\vspace{-15pt}
\caption{Numerical instrumentally resolved non-equilibrium metrics $\hat{I}_{\lambda}^{\,\text{ne},\text{Red}}(\lambda)$, obtained with $\Lambda^{\text{VUV}}=1$ (solid black lines), $\Lambda^{\text{VUV}}=0.1$ (solid blue lines), and $\Lambda^{\text{VUV}}=0.01$ (solid red lines), as well as the respective experimental instrumentally resolved non-equilibrium metrics $\hat{I}_{\lambda}^{\,\text{ne},\text{Red},\text{exp}}(\lambda)$ (dotted black lines).}
\label{fig:Results_I_ll_ne_inst_Red_1D_LL}
\end{figure}

\begin{figure}[H]
\centering
\centerline{\includegraphics[scale=1]{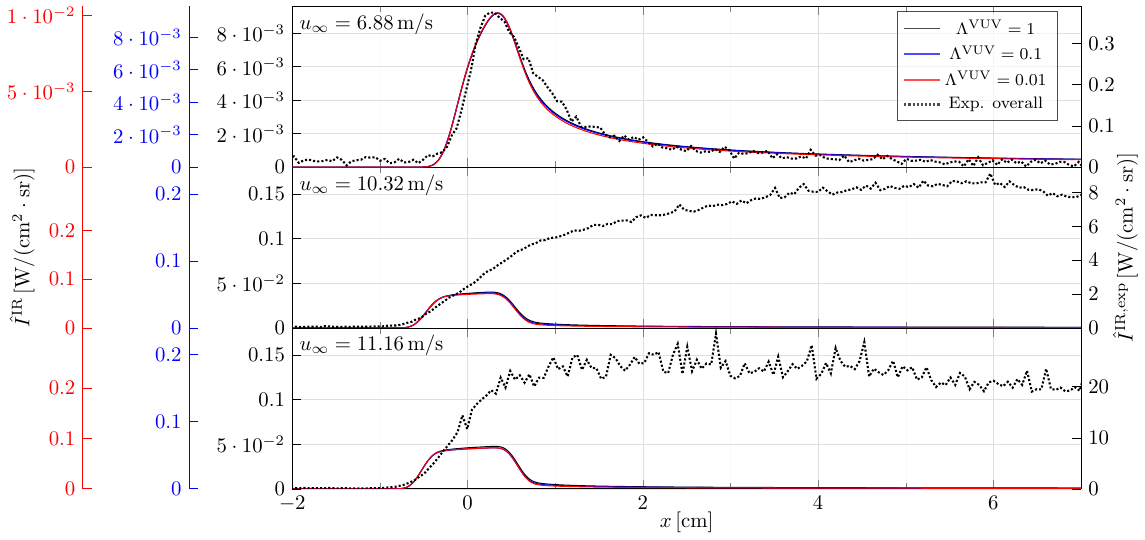}}
\vspace{-15pt}
\caption{Numerical instrumentally resolved radiative intensities $\hat{I}^{\text{IR}}(x)$, obtained with $\Lambda^{\text{VUV}}=1$ (solid black lines), $\Lambda^{\text{VUV}}=0.1$ (solid blue lines), and $\Lambda^{\text{VUV}}=0.01$ (solid red lines), as well as the respective experimental instrumentally resolved radiative intensities $\hat{I}^{\text{IR},\text{exp}}(x)$ (dotted black lines).}
\label{fig:Results_I_inst_IR_1D_LL}
\end{figure}
\begin{figure}[H]
\centering
\centerline{\includegraphics[scale=1]{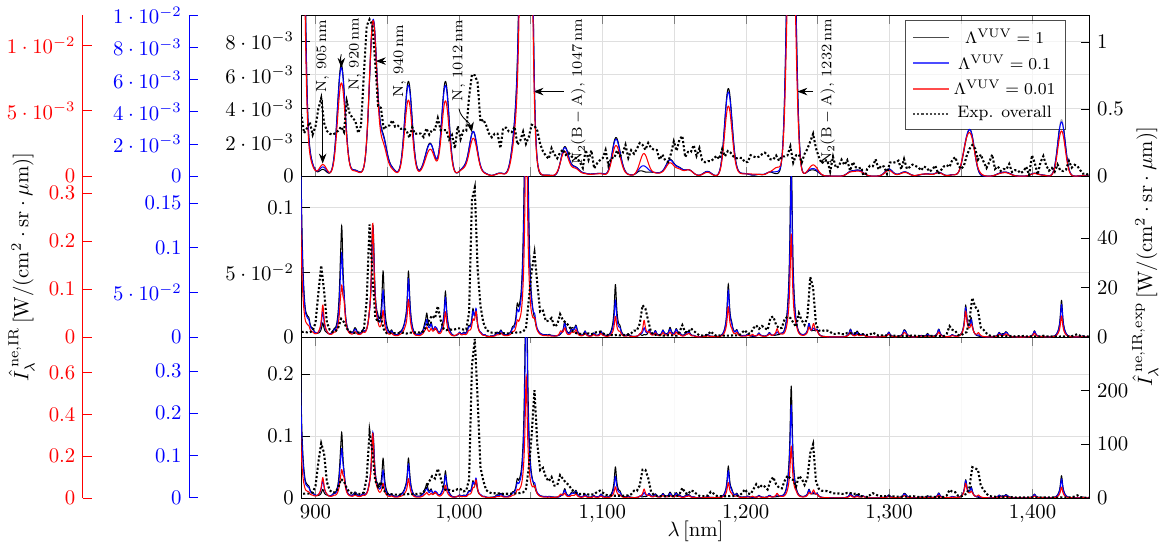}}
\vspace{-15pt}
\caption{Numerical instrumentally resolved non-equilibrium metrics $\hat{I}_{\lambda}^{\,\text{ne},\text{IR}}(\lambda)$, obtained with $\Lambda^{\text{VUV}}=1$ (solid black lines), $\Lambda^{\text{VUV}}=0.1$ (solid blue lines), and $\Lambda^{\text{VUV}}=0.01$ (solid red lines), as well as the respective experimental instrumentally resolved non-equilibrium metrics $\hat{I}_{\lambda}^{\,\text{ne},\text{IR},\text{exp}}(\lambda)$ (dotted black lines).}
\label{fig:Results_I_ll_ne_inst_IR_1D_LL}
\end{figure}

\subsection{Dependence on the dissociation rates of \ch{N2}}
\label{section:appendixVD_dependence}

\begin{figure}[H]
\centering
\centerline{\includegraphics[scale=1]{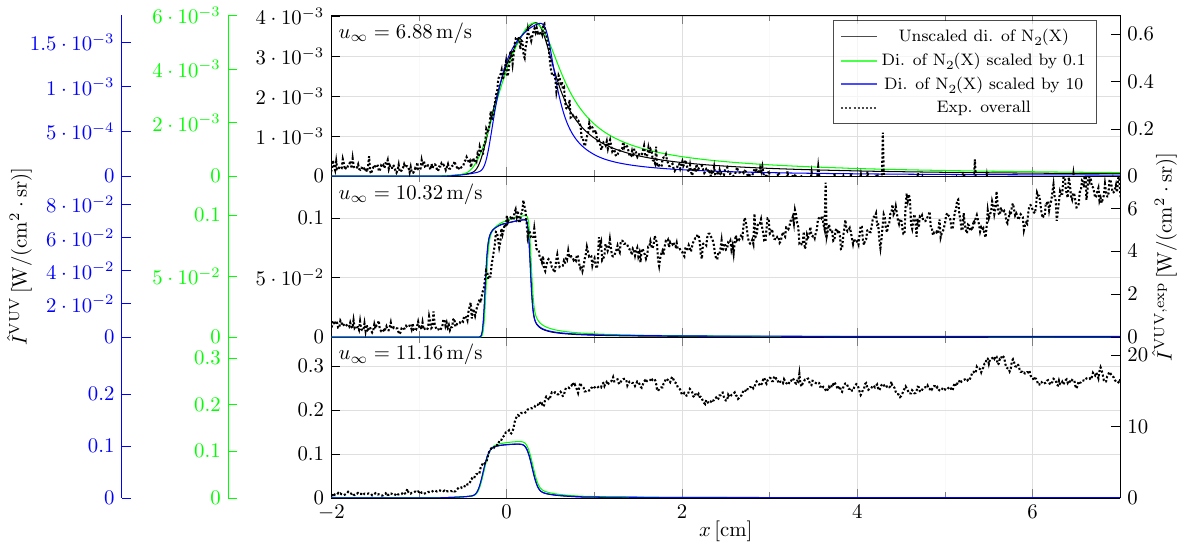}}
\vspace{-15pt}
\caption{Numerical instrumentally resolved radiative intensities $\hat{I}^{\text{VUV}}(x)$, obtained with $\Lambda^{\text{VUV}}=0.01$, and unscaled dissociation of $\ch{N2}(\text{X}{}^1\Sigma_\text{g}^+)$ (solid black lines), and scaled by $0.1$ (solid green lines), and by $10$ (solid blue lines), as well as the respective experimental instrumentally resolved radiative intensities $\hat{I}^{\text{VUV},\text{exp}}(x)$ (dotted black lines).}
\label{fig:Results_I_inst_VUV_1D_VD}
\end{figure}
\begin{figure}[H]
\centering
\centerline{\includegraphics[scale=1]{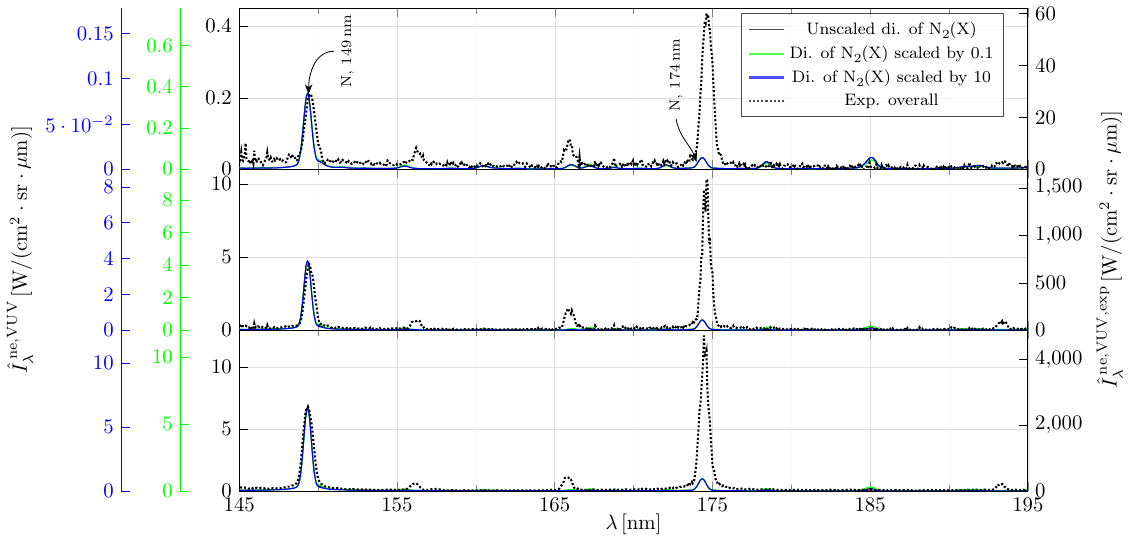}}
\vspace{-15pt}
\caption{Numerical instrumentally resolved non-equilibrium metrics $\hat{I}_{\lambda}^{\,\text{ne},\text{VUV}}(\lambda)$, obtained with $\Lambda^{\text{VUV}}=0.01$, and unscaled dissociation of $\ch{N2}(\text{X}{}^1\Sigma_\text{g}^+)$ (solid black lines), and scaled by $0.1$ (solid green lines), and by $10$ (solid blue lines), as well as the respective experimental instrumentally resolved non-equilibrium metrics $\hat{I}_{\lambda}^{\,\text{ne},\text{VUV},\text{exp}}(\lambda)$ (dotted black lines).}
\label{fig:Results_I_ll_ne_inst_VUV_1D_VD}
\end{figure}

\begin{figure}[H]
\centering
\centerline{\includegraphics[scale=1]{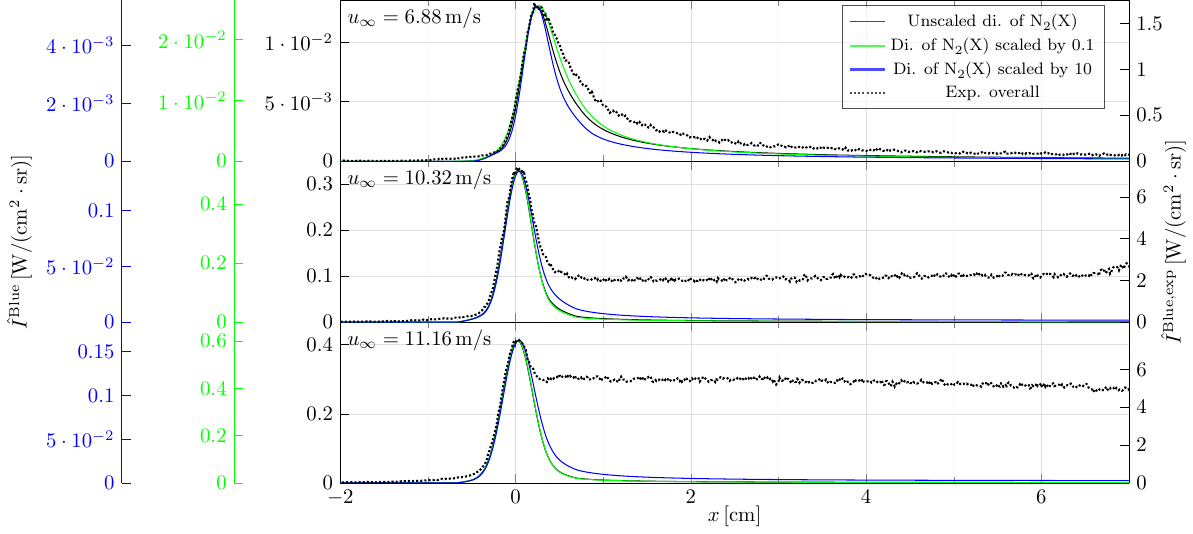}}
\vspace{-15pt}
\caption{Numerical instrumentally resolved radiative intensities $\hat{I}^{\text{Blue}}(x)$, obtained with $\Lambda^{\text{VUV}}=0.01$, and unscaled dissociation of $\ch{N2}(\text{X}{}^1\Sigma_\text{g}^+)$ (solid black lines), and scaled by $0.1$ (solid green lines), and by $10$ (solid blue lines), as well as the respective experimental instrumentally resolved radiative intensities $\hat{I}^{\text{Blue},\text{exp}}(x)$ (dotted black lines).}
\label{fig:Results_I_inst_Blue_1D_VD}
\end{figure}
\begin{figure}[H]
\centering
\centerline{\includegraphics[scale=1]{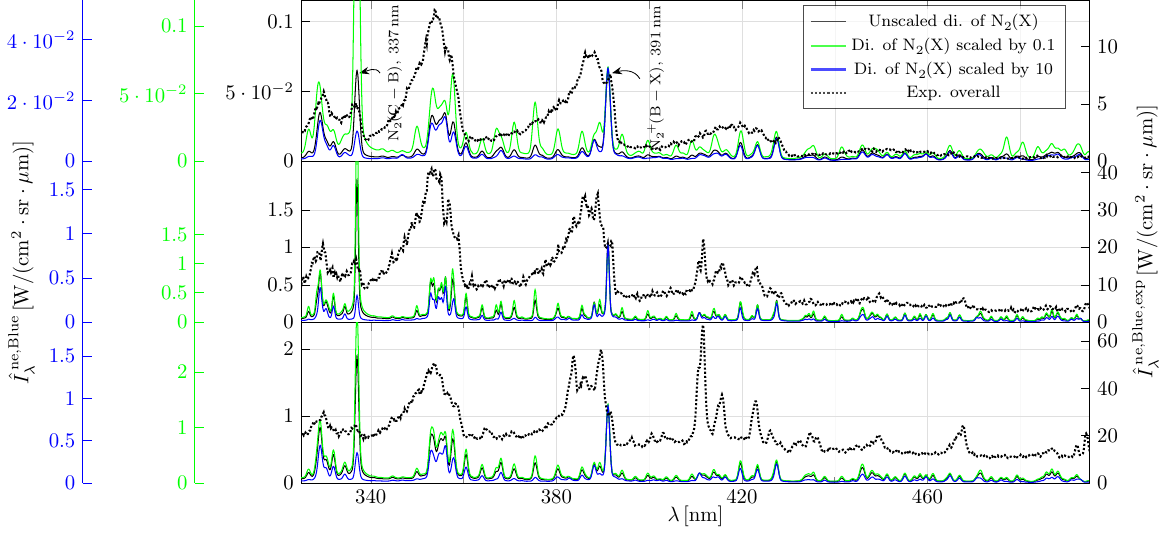}}
\vspace{-15pt}
\caption{Numerical instrumentally resolved non-equilibrium metrics $\hat{I}_{\lambda}^{\,\text{ne},\text{Blue}}(\lambda)$, obtained with $\Lambda^{\text{VUV}}=0.01$, and unscaled dissociation of $\ch{N2}(\text{X}{}^1\Sigma_\text{g}^+)$ (solid black lines), and scaled by $0.1$ (solid green lines), and by $10$ (solid blue lines), as well as the respective experimental instrumentally resolved non-equilibrium metrics $\hat{I}_{\lambda}^{\,\text{ne},\text{Blue},\text{exp}}(\lambda)$ (dotted black lines).}
\label{fig:Results_I_ll_ne_inst_Blue_1D_VD}
\end{figure}

\begin{figure}[H]
\centering
\centerline{\includegraphics[scale=1]{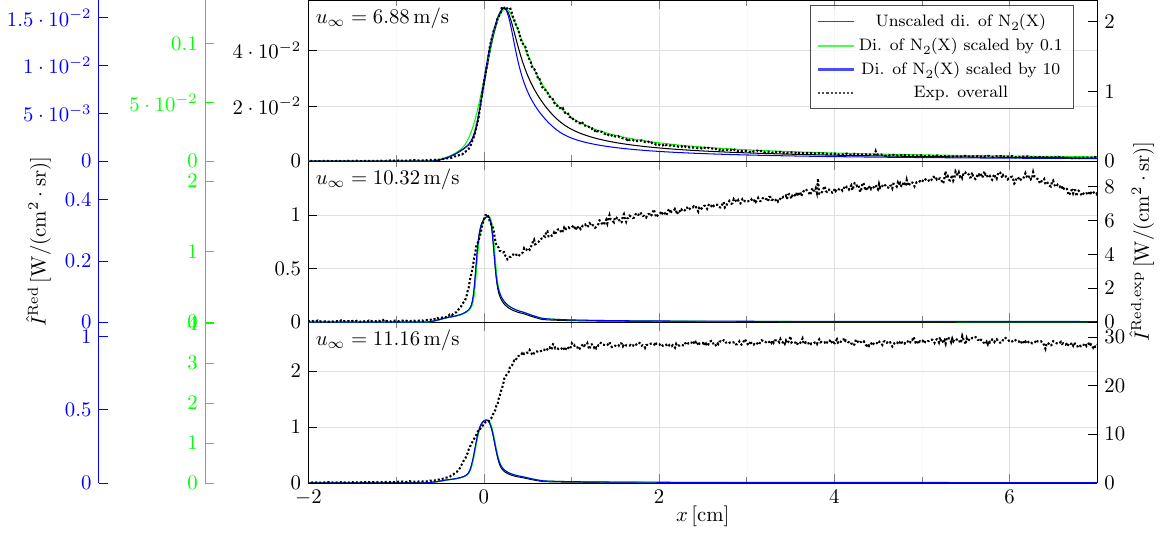}}
\vspace{-15pt}
\caption{Numerical instrumentally resolved radiative intensities $\hat{I}^{\text{Red}}(x)$, obtained with $\Lambda^{\text{VUV}}=0.01$, and unscaled dissociation of $\ch{N2}(\text{X}{}^1\Sigma_\text{g}^+)$ (solid black lines), and scaled by $0.1$ (solid green lines), and by $10$ (solid blue lines), as well as the respective experimental instrumentally resolved radiative intensities $\hat{I}^{\text{Red},\text{exp}}(x)$ (dotted black lines).}
\label{fig:Results_I_inst_Red_1D_VD}
\end{figure}
\begin{figure}[H]
\centering
\centerline{\includegraphics[scale=1]{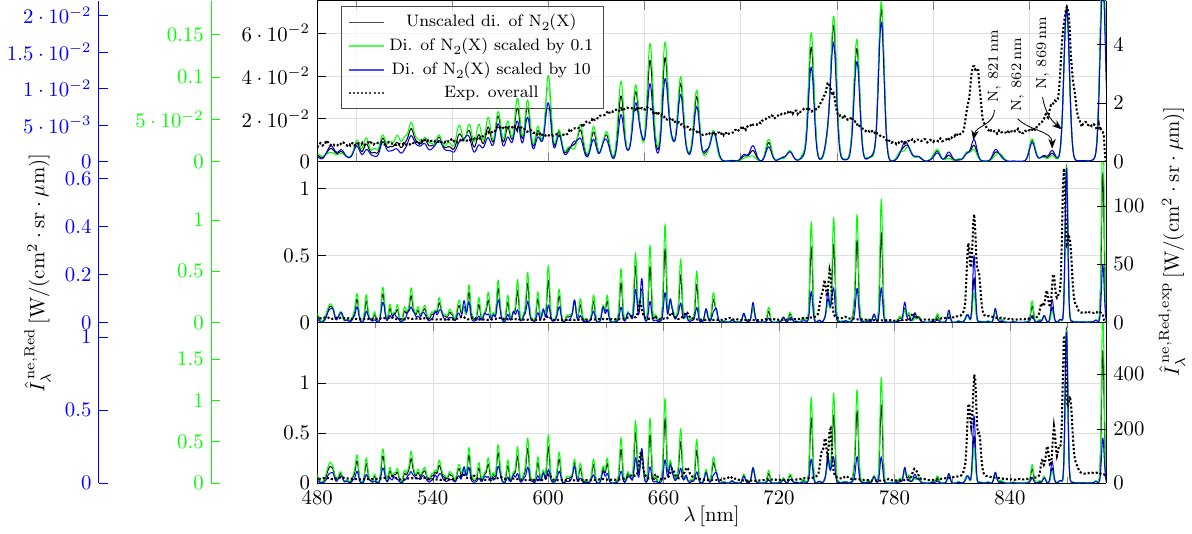}}
\vspace{-15pt}
\caption{Numerical instrumentally resolved non-equilibrium metrics $\hat{I}_{\lambda}^{\,\text{ne},\text{Red}}(\lambda)$, obtained with $\Lambda^{\text{VUV}}=0.01$, and unscaled dissociation of $\ch{N2}(\text{X}{}^1\Sigma_\text{g}^+)$ (solid black lines), and scaled by $0.1$ (solid green lines), and by $10$ (solid blue lines), as well as the respective experimental instrumentally resolved non-equilibrium metrics $\hat{I}_{\lambda}^{\,\text{ne},\text{Red},\text{exp}}(\lambda)$ (dotted black lines).}
\label{fig:Results_I_ll_ne_inst_Red_1D_VD}
\end{figure}

\begin{figure}[H]
\centering
\centerline{\includegraphics[scale=1]{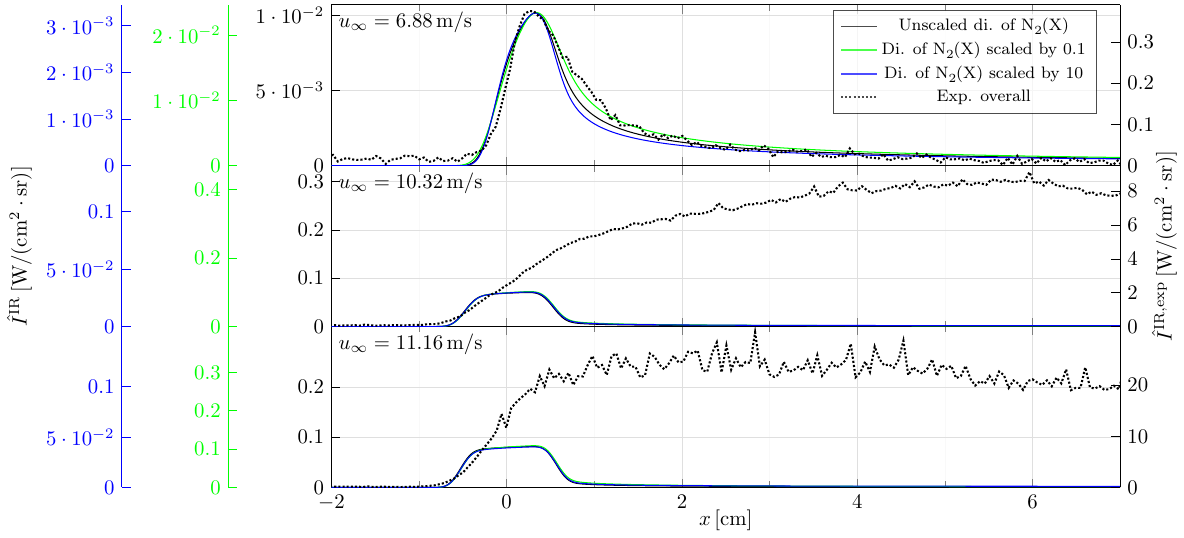}}
\vspace{-15pt}
\caption{Numerical instrumentally resolved radiative intensities $\hat{I}^{\text{IR}}(x)$, obtained with $\Lambda^{\text{VUV}}=0.01$, and unscaled dissociation of $\ch{N2}(\text{X}{}^1\Sigma_\text{g}^+)$ (solid black lines), and scaled by $0.1$ (solid green lines), and by $10$ (solid blue lines), as well as the respective experimental instrumentally resolved radiative intensities $\hat{I}^{\text{IR},\text{exp}}(x)$ (dotted black lines).}
\label{fig:Results_I_inst_IR_1D_VD}
\end{figure}
\begin{figure}[H]
\centering
\centerline{\includegraphics[scale=1]{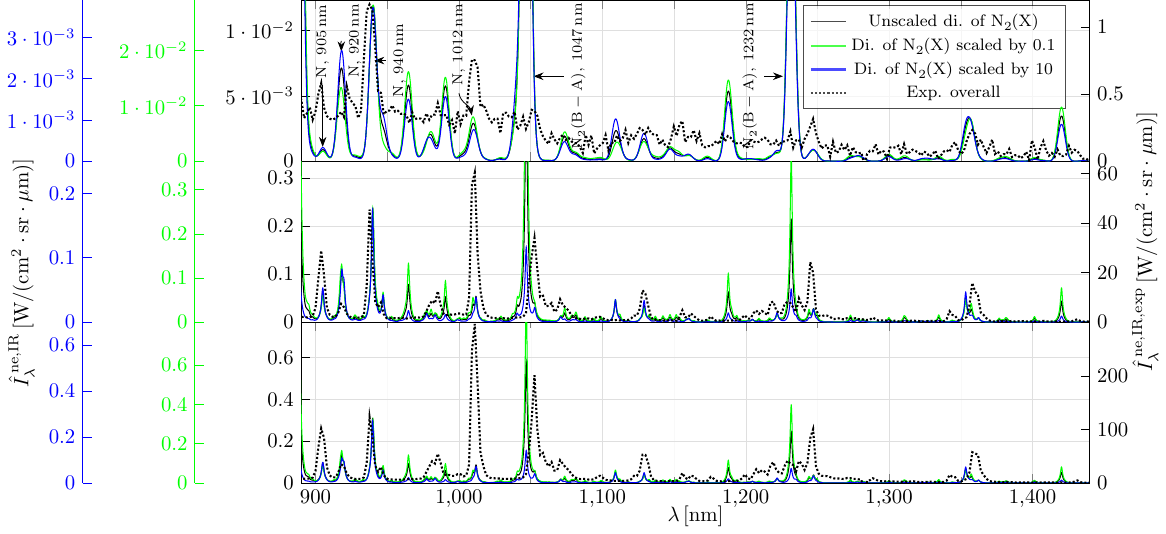}}
\vspace{-15pt}
\caption{Numerical instrumentally resolved non-equilibrium metrics $\hat{I}_{\lambda}^{\,\text{ne},\text{IR}}(\lambda)$, obtained with $\Lambda^{\text{VUV}}=0.01$, and unscaled dissociation of $\ch{N2}(\text{X}{}^1\Sigma_\text{g}^+)$ (solid black lines), and scaled by $0.1$ (solid green lines), and by $10$ (solid blue lines), as well as the respective experimental instrumentally resolved non-equilibrium metrics $\hat{I}_{\lambda}^{\,\text{ne},\text{IR},\text{exp}}(\lambda)$ (dotted black lines).}
\label{fig:Results_I_ll_ne_inst_IR_1D_VD}
\end{figure}

\subsection{Dependence on the excitation rates of \ch{N}}
\label{section:appendix_Excitation_of_N}

\begin{figure}[H]
\centering
\centerline{\includegraphics[scale=1]{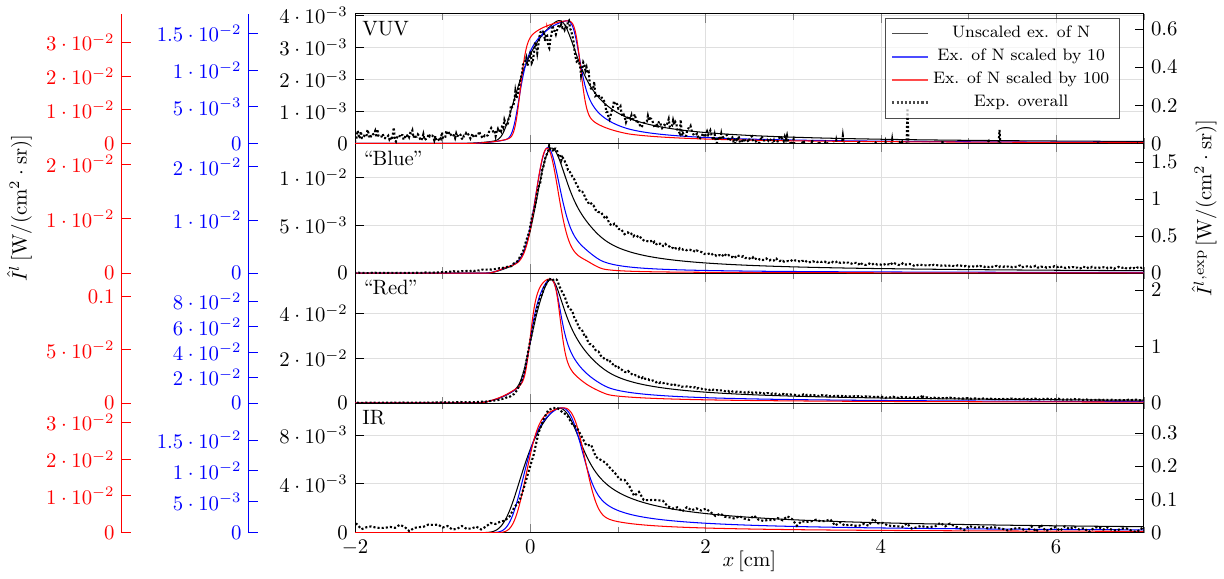}}
\vspace{-15pt}
\caption{Numerical instrumentally resolved radiative intensities $\hat{I}^{l}(x)$  obtained with $\Lambda^{\text{VUV}}=0.01$, and unscaled excitation of \ch{N} (solid black lines), and scaled by $10$ (solid blue lines), and by $100$ (solid red lines), as well as the respective experimental instrumentally resolved radiative intensities $\hat{I}^{l,\text{exp}}(x)$ (dotted black lines), for the case of the low speed shot.}
\label{fig:Results_I_inst_1D_LL_E_N}
\end{figure}
\begin{figure}[H]
\captionsetup[subfloat]{captionskip=-15pt}
\centering
\subfloat[][For the case $l=\text{VUV}$.]{
\centerline{\includegraphics[scale=1]{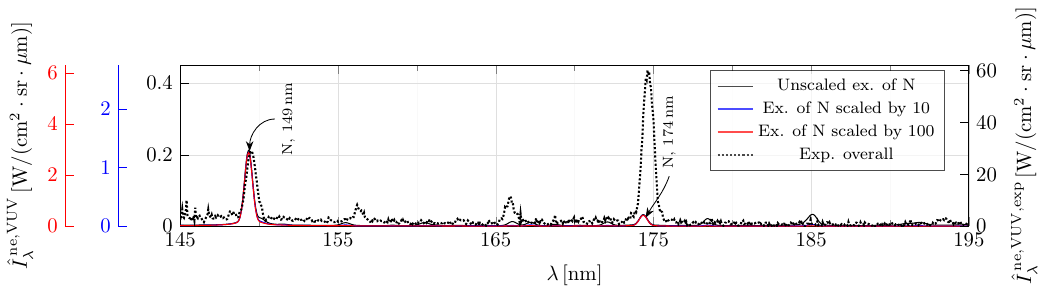}}
\label{fig:Results_I_ll_ne_inst_VUV_1D_LL_E_N}
}

\vspace{-33pt}

\subfloat[][For the case $l=\text{``Blue''}$.]{
\centering
\hspace{-10pt}\centerline{\includegraphics[scale=1]{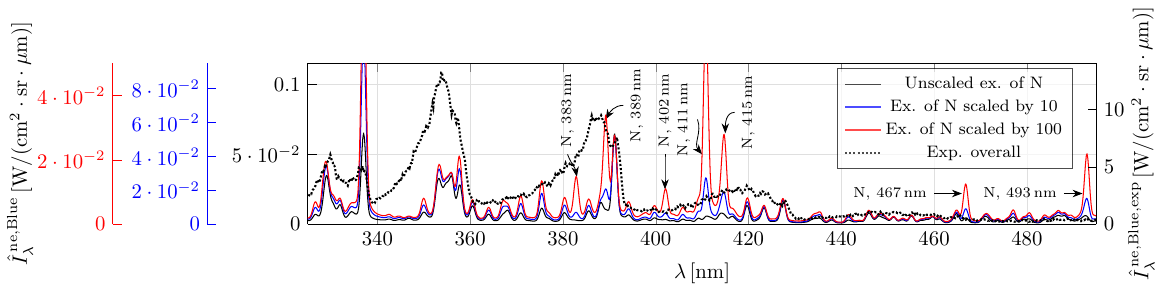}}
\label{fig:Results_I_ll_ne_inst_Blue_1D_LL_E_N}
}

\vspace{-15pt}

\subfloat[][For the case $l=\text{``Red''}$.]{
\centering
\hspace{-10pt}\centerline{\includegraphics[scale=1]{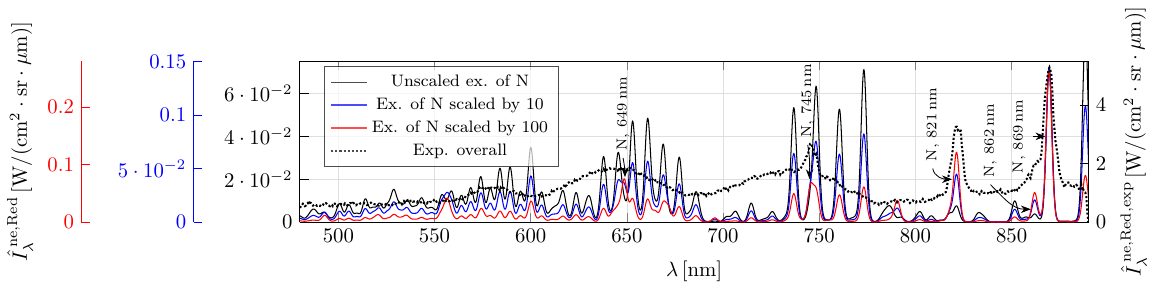}}
\label{fig:Results_I_ll_ne_inst_Red_1D_LL_E_N}
}

\vspace{-33pt}

\subfloat[][For the case $l=\text{IR}$.]{
\centering
\hspace{-10pt}\centerline{\includegraphics[scale=1]{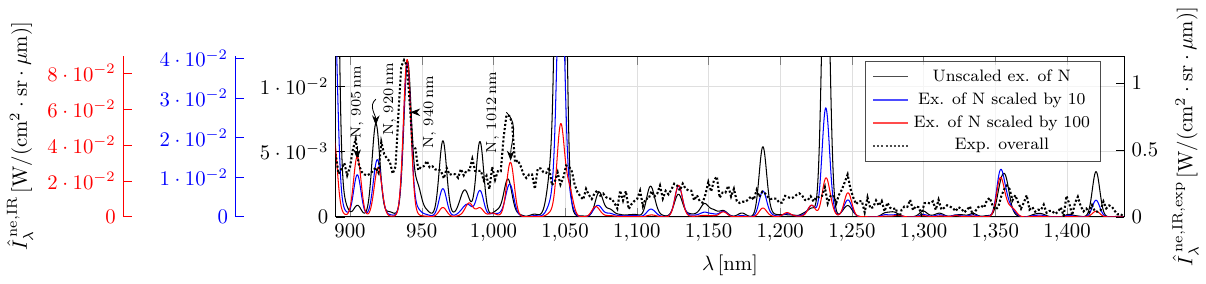}}
\label{fig:Results_I_ll_ne_inst_IR_1D_LL_E_N}
}

\vspace{-10pt}

\caption{Numerical instrumentally resolved non-equilibrium metrics $\hat{I}_{\lambda}^{\,\text{ne},l}(\lambda)$, obtained with $\Lambda^{\text{VUV}}=0.01$, and unscaled excitation of \ch{N} (solid black lines), and scaled by $10$ (solid blue lines), and by $100$ (solid red lines), as well as the respective experimental instrumentally resolved non-equilibrium metrics $\hat{I}_{\lambda}^{\,\text{ne},l,\text{exp}}(\lambda)$ (dotted black lines), for the case of the low speed shot.}
\label{fig:Results_I_ll_ne_inst_1D_LL_E_N}
\end{figure} 

\pagebreak

\end{document}